%% file: main.tex
\documentclass[11pt,a4paper]{article}
\usepackage{jheppub}
\usepackage[T1]{fontenc}
\input{figures/fonts/figure_fonts.tex}
\usepackage{xcolor}
\usepackage{tikz}
\usetikzlibrary{calc,arrows.meta}
\usepackage{array,tabularx,booktabs}
\hypersetup{hidelinks,
  pdftitle={Melonic ladder holography with a single free bulk field in AdS(d+1)},
  pdfauthor={Kok Hong Thong}}
\renewcommand{\afterTocRuleSpace}{\clearpage}
\renewcommand{\acknowledgments}{\section*{Acknowledgements}}
\title{\boldmath Melonic ladder holography with a single free bulk field in AdS$_{d+1}$}
\author{Kok Hong Thong}
\affiliation{Centre for Theoretical Physics and Astronomy, Queen Mary University of London,\\ Mile End Road, London E1 4NS, United Kingdom}
\emailAdd{k.h.thong@qmul.ac.uk}
\abstract{We derive a cylinder from the pair Casimir of a CFT$_d$: at zero centre momentum it depends only on relative dilations and rotations. Relative-coordinate radial quantisation then encodes exchanged family labels as cylinder energies. For a melonic CFT$_d$ in the normalisable and stable range, the reduced ladder sum is the Robin boundary-to-boundary propagator of a single free scalar field in rigid AdS$_{d+1}$, with the cylinder as its conformal boundary. The Robin boundary condition and bulk regularity (or parity) determine the holographic radial extension of the boundary data. A nonzero Robin boundary coefficient in global AdS breaks $2d$ AdS generators, yet the scalar's boundary propagator still encodes the conformal data of the melonic ladder. We demonstrate the correspondence for the Sachdev--Ye--Kitaev and the $d=2$ disorder-free long-range Amit--Roginsky melonic models, which correspond to a single free local scalar in AdS$_2$ and AdS$_3$, respectively.}
\begin{document}
\maketitle
\flushbottom
\input{sections/01_introduction}
\input{sections/02_relative_cylinder}
\input{sections/03_robin_ladders}

\input{sections/05_examples}

\input{sections/06_discussion}
\acknowledgments
The author is grateful to D. Vegh for many helpful discussions and comments on a draft of this paper. 
The author also acknowledges the hospitality of BUSSTEPP 2026 in Cambridge, from which this work greatly benefited. 
The author is supported by a Science and Technology Facilities Council
(STFC) studentship. 
We acknowledge the use of AI tools for assistance in making figures and editing the manuscript.
No new data were generated or analysed during this study.
\appendix
\input{appendices/A_reduction}
\input{appendices/B_scalar}
\input{appendices/C_syk}
\input{appendices/D_ar}
\input{appendices/E_dgjs}
\input{appendices/F_folded_string}
\input{references}
\end{document}

%% file: figures/fonts/figure_fonts.tex
\ifdefined\FigureSans\else%
  \IfFileExists{t1lmss.fd}{%
    \def\MelonicFigureFamily{lmss}%
  }{%
    \def\MelonicFigureFamily{cmss}%
  }%
  \newcommand{\FigureSans}{%
    \fontencoding{T1}\fontfamily{\MelonicFigureFamily}\selectfont}%
\fi%

%% file: sections/01_introduction.tex
\section{Introduction}
\label{sec:intro}
Melonic large-$N$ CFTs are a class of CFTs that are exactly solvable, even in the strong-coupling limit.
The propagators are dressed by an infinite sum of iterated watermelon-like diagrams (figure~\ref{fig:melons}) and the leading connected four-point functions are sums of ladder diagrams constructed rung by rung~\cite{MS16,Ben20}; higher-point functions are obtained by joining these building blocks~\cite{GR17all}.
Due to their tractability, melonic models, such as Sachdev--Ye--Kitaev (SYK), have been studied in detail.
SYK ($\sim$CFT$_1$) is especially relevant for holography, black holes and chaos~\cite{MS16,Rosenhaus18,Sarosi17}.
A CFT in $d$ dimensions organises a four-point function into conformal families with fixed Casimir eigenvalues; each family includes a primary and its descendants~\cite{SimmonsDuffin16}.
In the conventional large-$N$ dictionary, each single-trace primary operator corresponds to a single bulk field, so an infinite tower of fields is usually required in the AdS$_{d+1}$ bulk~\cite{JSY16,GR17cubic,Sarosi17}.
Here we ask whether a single local free AdS$_{d+1}$ scalar can encode the conformal multiplet data of a conformal melonic ladder. 

\input{figures/melons}

Such a single-field description is far from obvious in the conventional bulk
dictionary: Maldacena and Stanford interpreted the SYK bilinear tower with $O(1)$ separation as
suggestive of low-tension strings, with a string length of order the AdS
radius~\cite{MS16}. A scalar with a nontrivial potential in a spacetime
conformal to $\mathrm{AdS}_2\times S^1/\mathbb{Z}_2$ reproduces the SYK bilinear
spectrum and the bilocal propagator~\cite{DGJS17a,DJS17,DGJS17b}, but its Kaluza--Klein
reduction gives an infinite tower of $\mathrm{AdS}_2$ fields, rather than a
single field. 
Vegh's folded string construction reproduces the bilinear spectrum
in $\mathrm{AdS}_2$; however, the AdS spacetime has imaginary radius squared~\cite{Vegh24,Vegh25}.
The folded string approach can be reduced to a quantum mechanical problem on an interval with fractional-power Robin boundary conditions~\cite{TV26}.
These constructions motivate the search for a local free scalar in $\mathrm{AdS}_2$, perhaps with Robin boundary conditions, that can encode the SYK melonic ladder.
Since fields with Robin boundary conditions break AdS generators, what physical principles could make such an identification possible, and could they extend to other melonic and related tensor models~\cite{Ben20,Witten16,KT16,GRline,GKT17,BGH19,GKPPP18,BDHS20,MSW17,Popov19,BD20,GNYP23,FTW24,BLM26}?

Using the principles of radial quantisation and integrating out the bilocal centres, we derive a cylinder $\mathbb{R}\times S^{d-1}$ from the pair Casimir, whose cylinder energies encode the exchanged CFT dimensions through $\omega=h-d/2$. 
Surprisingly, within the normalisable and stable range, conformal melonic ladders select a single local free AdS$_{d+1}$ scalar whose Robin boundary condition carries the ladder coupling. 
The conformal boundary of this AdS$_{d+1}$ is the derived cylinder; the Robin scalar field breaks $2d$ of the AdS$_{d+1}$ generators, and the remaining symmetries are exactly those of the cylinder. The procedure suggests an interesting duality: conformal melonic ladders correspond to Robin boundary-to-boundary propagators of a single bulk scalar field, where each rung of the ladder corresponds to a Neumann boundary-to-boundary propagator. 
The scaling dimension of the constituent field of the melonic CFT$_d$, such as the SYK fermion, determines the mass of the AdS$_{d+1}$ scalar, while the bilinear spectrum corresponds to a shifted energy spectrum of the scalar.
Bulk regularity for $d>1$ (or parity for $d=1$) and normalisation fix the radial dependence of each bulk mode. 

We illustrate this with two examples of melonic CFTs: (i) SYK ($d=1$), with disorder, and (ii) the long-range Amit--Roginsky (AR) model with $d=2$, without disorder. 
These correspond to a single free local scalar in rigid AdS$_2$ and AdS$_3$, respectively. 
We caution that this is not an equivalence between the two full theories.
For example, higher-point functions and finite-$N$ effects are not identified.
Stability of the scalar is also a separate criterion, which we discuss in section~\ref{sec:examples}.

%% file: figures/melons.tex
\begin{figure}[htbp]
\centering
\resizebox{.48\textwidth}{!}{\input{figures/artwork/melons.tex}}
\caption{A Feynman diagram of the propagator with 20 quartic melons. The full propagator contains infinitely many iterated melons.}
\label{fig:melons}
\end{figure}
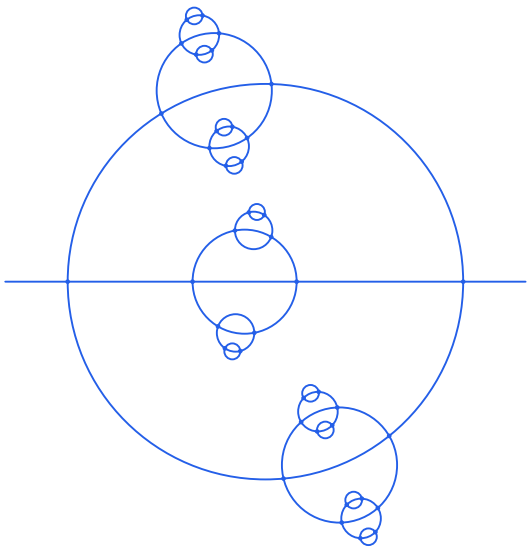

%% file: figures/artwork/melons.tex
\begingroup
\definecolor{MelonicBlue}{RGB}{36,95,232}
\definecolor{MelonicBackground}{RGB}{238,243,255}
\begin{tikzpicture}[x=1cm,y=1cm,
  propagator/.style={draw=MelonicBlue,line width=.78pt,
                     line cap=round,line join=round}]
\draw[propagator] (-3.75,0) -- (3.75,0);

\draw[propagator] (0.00000000,0.00000000) circle[radius=2.85000cm];
\draw[propagator] (-0.73763428,2.75288860) circle[radius=0.83000cm];
\draw[propagator] (-0.30000000,0.00000000) circle[radius=0.75000cm];
\draw[propagator] (1.06762879,-2.64247399) circle[radius=0.83000cm];
\draw[propagator] (-0.95245409,3.55460704) circle[radius=0.28500cm];
\draw[propagator] (-0.52281447,1.95117017) circle[radius=0.28500cm];
\draw[propagator] (-0.16976387,0.73860581) circle[radius=0.27000cm];
\draw[propagator] (-0.43023613,-0.73860581) circle[radius=0.27000cm];
\draw[propagator] (1.37855226,-3.41203658) circle[radius=0.28500cm];
\draw[propagator] (0.75670532,-1.87291139) circle[radius=0.28500cm];
\draw[propagator] (-1.02621751,3.82989590) circle[radius=0.12000cm];
\draw[propagator] (-0.87869066,3.27931818) circle[radius=0.12000cm];
\draw[propagator] (-0.44905104,1.67588131) circle[radius=0.12000cm];
\draw[propagator] (-0.59657790,2.22645903) circle[radius=0.12000cm];
\draw[propagator] (1.48531514,-3.67628398) circle[radius=0.12000cm];
\draw[propagator] (1.27178938,-3.14778919) circle[radius=0.12000cm];
\draw[propagator] (0.64994244,-1.60866399) circle[radius=0.12000cm];
\draw[propagator] (0.86346820,-2.13715878) circle[radius=0.12000cm];
\draw[propagator] (-0.12287886,1.00450391) circle[radius=0.11500cm];
\draw[propagator] (-0.47712114,-1.00450391) circle[radius=0.11500cm];
\fill[MelonicBlue] (-2.85000000,0.00000000) circle[radius=.96pt];
\fill[MelonicBlue] (2.85000000,0.00000000) circle[radius=.96pt];
\fill[MelonicBlue] (-1.49952680,2.42361700) circle[radius=.96pt];
\fill[MelonicBlue] (0.08681980,2.84867729) circle[radius=.96pt];
\fill[MelonicBlue] (-1.05000000,0.00000000) circle[radius=.96pt];
\fill[MelonicBlue] (0.45000000,0.00000000) circle[radius=.96pt];
\fill[MelonicBlue] (1.78371417,-2.22280538) circle[radius=.96pt];
\fill[MelonicBlue] (0.26099377,-2.83802436) circle[radius=.96pt];
\fill[MelonicBlue] (-1.21099114,3.43467543) circle[radius=.96pt];
\fill[MelonicBlue] (-0.66858862,3.58001175) circle[radius=.96pt];
\fill[MelonicBlue] (-0.26427741,2.07110178) circle[radius=.96pt];
\fill[MelonicBlue] (-0.80667993,1.92576546) circle[radius=.96pt];
\fill[MelonicBlue] (-0.43975824,0.73686337) circle[radius=.96pt];
\fill[MelonicBlue] (0.08335191,0.64462494) circle[radius=.96pt];
\fill[MelonicBlue] (-0.16024176,-0.73686337) circle[radius=.96pt];
\fill[MelonicBlue] (-0.68335191,-0.64462494) circle[radius=.96pt];
\fill[MelonicBlue] (1.62054624,-3.26149119) circle[radius=.96pt];
\fill[MelonicBlue] (1.09989875,-3.47184643) circle[radius=.96pt];
\fill[MelonicBlue] (0.51471134,-2.02345678) circle[radius=.96pt];
\fill[MelonicBlue] (1.03535884,-1.81310154) circle[radius=.96pt];
\fill[MelonicBlue] (-1.13299225,3.77513135) circle[radius=.96pt];
\fill[MelonicBlue] (-0.90636561,3.83585578) circle[radius=.96pt];
\fill[MelonicBlue] (-0.77191592,3.33408273) circle[radius=.96pt];
\fill[MelonicBlue] (-0.99854257,3.27335830) circle[radius=.96pt];
\fill[MelonicBlue] (-0.34227631,1.73064586) circle[radius=.96pt];
\fill[MelonicBlue] (-0.56890295,1.66992143) circle[radius=.96pt];
\fill[MelonicBlue] (-0.70335263,2.17169448) circle[radius=.96pt];
\fill[MelonicBlue] (-0.47672599,2.23241891) circle[radius=.96pt];
\fill[MelonicBlue] (1.58461988,-3.60891507) circle[radius=.96pt];
\fill[MelonicBlue] (1.36708292,-3.69680571) circle[radius=.96pt];
\fill[MelonicBlue] (1.17248465,-3.21515810) circle[radius=.96pt];
\fill[MelonicBlue] (1.39002161,-3.12726746) circle[radius=.96pt];
\fill[MelonicBlue] (0.55063771,-1.67603290) circle[radius=.96pt];
\fill[MelonicBlue] (0.76817466,-1.58814226) circle[radius=.96pt];
\fill[MelonicBlue] (0.96277293,-2.06978987) circle[radius=.96pt];
\fill[MelonicBlue] (0.74523597,-2.15768051) circle[radius=.96pt];
\fill[MelonicBlue] (-0.23778653,0.99989668) circle[radius=.96pt];
\fill[MelonicBlue] (-0.01647673,0.96087379) circle[radius=.96pt];
\fill[MelonicBlue] (-0.36221347,-0.99989668) circle[radius=.96pt];
\fill[MelonicBlue] (-0.58352327,-0.96087379) circle[radius=.96pt];
\end{tikzpicture}
\endgroup

%% file: sections/02_relative_cylinder.tex
\section{The CFT pair and its relative cylinder}
\label{sec:cylinder}
We take scalar constituent fields $\phi$ of dimension $\Delta_\phi$ in a CFT$_d$; the fermionic $d=1$ case is SYK. Write $x_{ij}\equiv x_i-x_j$, $X\equiv(x_1+x_2)/2$, and $x\equiv x_{12}$. On bilocal functions $f(X,x)$, the differential pair momentum and dilation generators are given by
\begin{equation}
P_{12}\equiv\partial_X,\qquad D_{12}\equiv X\cdot\partial_X+x\cdot\partial_x+2\Delta_\phi,
\label{eq:gens}
\end{equation}
where we have dropped the factor of $i$.
The pair Casimir is
\begin{equation}
\begin{split}
C_{12}&\equiv D_{12}(D_{12}-d)+\mathbf{L}_{12}^2-K_{12}\cdot P_{12}.
\end{split}
\label{eq:C12full}
\end{equation}
Here $\mathbf{L}_{12}$ and $K_{12}$ are the pair rotation and special conformal generators, respectively, with the operator ordering shown (appendix~\ref{app:A}). 
Define the relative dilation and angular momentum operators as
\begin{equation}
    D_{\rm rel}\equiv x\cdot\partial_x+2\Delta_\phi, \qquad \mathbf{L}^2\equiv-\nabla^2_{S^{d-1}},
\end{equation}
respectively. 
Select bilocals with zero centre (total) momentum, $P_{12}f=0$. 
For an exchanged family of dimension $h$ and spin $J$, the pair Casimir becomes
\begin{equation}
\begin{split}
C_{12}|_{P_{12}=0}&=C_{\rm rel}\equiv D_{\rm rel}(D_{\rm rel}-d)+\mathbf{L}^2.
\end{split}
\label{eq:Crel}
\end{equation}
Its eigenvalue is $h(h-d)+J(J+d-2)$.

Turn the punctured Euclidean relative space into a cylinder: $x\equiv e^\tau\mathbf{n}$, $\mathbf{n}\in S^{d-1}$, with $\mathbf{n}^2=1$. 
To remove the first derivative in $\tau$ from the Casimir, define $a\equiv2\Delta_\phi-d/2$ and $(Wf)(\tau,\mathbf{n})\equiv e^{a\tau}f(e^\tau\mathbf{n})$. This gives
\begin{equation}
\begin{split}
WD_{\rm rel}W^{-1}&=\partial_\tau+d/2,\\
WC_{\rm rel}W^{-1}&=\partial_\tau^2+\mathbf{L}^2-d^2/4.
\end{split}
\label{eq:Ccyl}
\end{equation}
Here $WC_{\rm rel}W^{-1}$ is the reduced pair conformal Casimir acting on the relative cylinder $\mathbb R\times S^{d-1}$ with coordinates $(\tau,\mathbf n)$.
This is a relative-coordinate radial quantisation (see section~\ref{sec:discussion} for related bilocal constructions~\cite{DGJS17b,PR16}).
A unit-normalised degree-$J$ harmonic $Y_J(\mathbf{n})$, with degeneracy suppressed, gives the mode
\begin{equation}
e^{-\omega\tau}Y_J(\mathbf{n}),\qquad \omega\equiv h-d/2.
\label{eq:mode}
\end{equation}
The shadow $d-h$ reverses the sign of $\omega$.
In $d=1$, $\mathbf{L}^2=0$ and the values $J=0,1$ label even and odd functions on $S^0$, respectively, not spin. 
In ordinary CFT radial quantisation, a primary of dimension $h$ and its level-$n$ descendants have energies $h+n$~\cite{PRER12}. 
The pair generators $P_{12}$ and $K_{12}$ raise and lower descendant levels within each family, so dilation eigenvalues alone do not identify conformal families. Here, each cylinder energy eigenvalue represents the Casimir eigenvalue of a conformal family. 

We implement the centre reduction by integration. 
Appendix~\ref{app:A} gives the conditions under which this integral preserves the pair Casimir.
The full pair-generator action does not survive under this reduction.

\input{figures/relative_cylinder}

This cylinder is the Euclidean continuation of $\partial$AdS$_{d+1}$ (figure~\ref{fig:cyl}). Can one local field realise its dynamics? Geometry alone fixes neither the mass nor the boundary condition of the field.

%% file: figures/relative_cylinder.tex
\begin{figure}[tbp]
\centering
\resizebox{\textwidth}{!}{\input{figures/artwork/relative_cylinder.tex}}
\caption{Cylinder from the relative-coordinate radial quantisation construction. Zero total momentum removes centre dependence; $x=e^\tau\mathbf{n}$ turns relative dilation into cylinder time translation and retains $\mathbf{n}\in S^{d-1}$.}
\label{fig:cyl}
\end{figure}
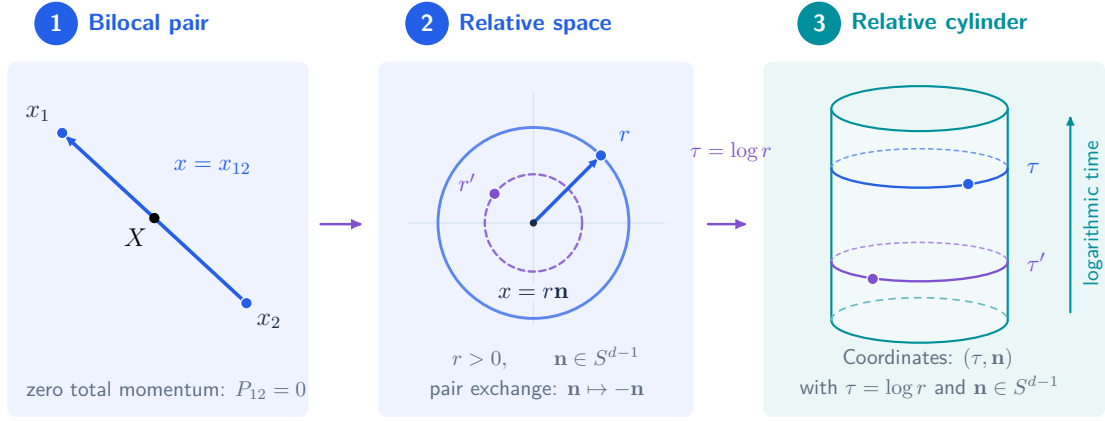

%% file: figures/artwork/relative_cylinder.tex
\begingroup%
\input{figures/fonts/figure_fonts.tex}%
\definecolor{melonBlue}{HTML}{245FE8}%
\definecolor{figureNavy}{HTML}{182B45}%
\definecolor{figureSlate}{HTML}{5D6E83}%
\definecolor{melonPanel}{HTML}{EEF3FF}%
\definecolor{scalarTeal}{HTML}{008F9C}%
\definecolor{boundaryOrange}{HTML}{E45A3D}%
\definecolor{mapPurple}{HTML}{8050CF}%
\definecolor{scalarPanel}{HTML}{EAF7F6}%
\definecolor{footerPanel}{HTML}{F7F9FC}%
\definecolor{figureTint1}{HTML}{D8E2EF}%
\definecolor{figureTint2}{HTML}{5B87EE}%
\definecolor{figureTint3}{HTML}{936AD6}%
\definecolor{figureTint4}{HTML}{F2F5FE}%
\definecolor{figureTint5}{HTML}{F2F9FA}%
\definecolor{figureTint6}{HTML}{66BCC4}%
\definecolor{figureTint7}{HTML}{7197F0}%
\definecolor{figureTint8}{HTML}{AC8DE0}%
\begin{tikzpicture}[x=1bp,y=-1bp,inner sep=0pt,outer sep=0pt]%
\path[use as bounding box] (0,0) rectangle (420,158);

\path[fill=melonBlue] (24.9104,12.4417) ellipse [x radius=7.9990bp,y radius=7.9990bp];
\path[fill=melonBlue] (161.1541,12.4417) ellipse [x radius=7.9990bp,y radius=7.9990bp];
\path[fill=scalarTeal] (305.2149,12.4418) ellipse [x radius=8.0235bp,y radius=8.0235bp];

\path[fill=melonPanel] (7.0426,153.8094)
    -- (7.0426,29.6529)
    .. controls (7.0426,27.9187) and (8.4484,26.5129) .. (10.1826,26.5129)
    -- (114.4610,26.5129)
    .. controls (116.1951,26.5129) and (117.6009,27.9187) .. (117.6009,29.6529)
    -- (117.6009,153.8094)
    .. controls (117.6009,155.5436) and (116.1951,156.9494) .. (114.4610,156.9494)
    -- (10.1826,156.9494)
    .. controls (8.4484,156.9494) and (7.0426,155.5436) .. (7.0426,153.8094);
\path[fill=melonPanel] (143.2860,153.8094)
    -- (143.2860,29.6529)
    .. controls (143.2860,27.9187) and (144.6918,26.5129) .. (146.4260,26.5129)
    -- (255.8417,26.5129)
    .. controls (257.5758,26.5129) and (258.9816,27.9187) .. (258.9816,29.6529)
    -- (258.9816,153.8094)
    .. controls (258.9816,155.5436) and (257.5758,156.9494) .. (255.8417,156.9494)
    -- (146.4260,156.9494)
    .. controls (144.6918,156.9494) and (143.2860,155.5436) .. (143.2860,153.8094);
\path[fill=scalarPanel] (284.6667,153.8094)
    -- (284.6667,29.6529)
    .. controls (284.6667,27.9187) and (286.0725,26.5129) .. (287.8067,26.5129)
    -- (407.0495,26.5129)
    .. controls (408.7837,26.5129) and (410.1895,27.9187) .. (410.1895,29.6529)
    -- (410.1895,153.8094)
    .. controls (410.1895,155.5436) and (408.7837,156.9494) .. (407.0495,156.9494)
    -- (287.8067,156.9494)
    .. controls (286.0725,156.9494) and (284.6667,155.5436) .. (284.6667,153.8094);

\path[draw=melonBlue,line width=1.33447bp,line cap=round] (94.8191,115.1830)
    -- (30.8347,56.0553);
\path[draw=melonBlue,line width=0.82012bp,line cap=round,fill=melonBlue] (28.8348,54.2072)
    .. controls (29.0210,54.4984) and (29.7645,55.8598) .. (30.0734,56.8792)
    -- (31.5960,55.2315)
    .. controls (30.5554,55.0039) and (29.1397,54.3700) .. (28.8348,54.2072);
\path[draw=white,line width=0.47099bp,line cap=round,fill=melonBlue] (27.1440,52.6449) ellipse [x radius=2.1095bp,y radius=2.1095bp];
\path[draw=white,line width=0.47099bp,line cap=round,fill=melonBlue] (94.8191,115.1830) ellipse [x radius=2.1095bp,y radius=2.1095bp];
\path[fill=melonPanel] (66.5661,61.0874) rectangle (98.0570,69.2175);

\path[draw=mapPurple,line width=0.90272bp,line cap=round] (122.0678,86.1474)
    -- (133.6866,86.1474);
\path[draw=mapPurple,line width=0.90272bp,line cap=round,fill=mapPurple] (136.5842,86.1474)
    .. controls (136.2283,86.0533) and (134.6642,85.5203) .. (133.6866,84.9403)
    -- (133.6866,87.3545)
    .. controls (134.6642,86.7744) and (136.2283,86.2414) .. (136.5842,86.1474);

\path[draw=figureTint1,line width=0.47099bp,line cap=round] (154.4536,85.7006)
    -- (248.2606,85.7006);
\path[draw=figureTint1,line width=0.47099bp,line cap=round] (200.2404,123.0000)
    -- (200.2404,37.6805);
\path[draw=figureTint2,line width=1.05973bp,line cap=round] (200.2404,85.7006) ellipse [x radius=35.0661bp,y radius=35.0661bp];
\path[draw=figureTint3,line width=0.86348bp,line cap=round,dash pattern=on 2.354959bp off 1.56997bp] (200.2404,85.7006) ellipse [x radius=17.8681bp,y radius=17.8681bp];
\path[fill=figureTint4] (200.2404,85.7006) ellipse [x radius=1.3400bp,y radius=1.3400bp];
\path[draw=melonBlue,line width=1.29521bp,line cap=round] (200.2404,85.7006)
    -- (221.7631,64.1779);
\path[draw=melonBlue,line width=0.72646bp,line cap=round,fill=melonBlue] (223.5347,62.4064)
    .. controls (223.2602,62.5670) and (221.9817,63.2011) .. (221.0333,63.4482)
    -- (222.4928,64.9077)
    .. controls (222.7399,63.9593) and (223.3740,62.6808) .. (223.5347,62.4064);
\path[draw=white,line width=0.47099bp,line cap=round,fill=melonBlue] (225.0323,60.9087) ellipse [x radius=2.1095bp,y radius=2.1095bp];
\path[draw=white,line width=0.47099bp,line cap=round,fill=mapPurple] (185.9460,74.9799) ellipse [x radius=2.1095bp,y radius=2.1095bp];
\path[fill=figureNavy] (200.2404,85.7006) ellipse [x radius=1.3345bp,y radius=1.3345bp];
\path[fill=melonPanel] (185.6796,107.9056) rectangle (214.8012,114.8658);

\path[draw=mapPurple,line width=0.90272bp,line cap=round] (264.3418,86.1474)
    -- (274.8441,86.1474);
\path[draw=mapPurple,line width=0.90272bp,line cap=round,fill=mapPurple] (277.7416,86.1474)
    .. controls (277.3858,86.0533) and (275.8217,85.5203) .. (274.8441,84.9403)
    -- (274.8441,87.3545)
    .. controls (275.8217,86.7744) and (277.3858,86.2414) .. (277.7416,86.1474);

\path[fill=figureTint5] (309.6818,121.4368)
    -- (309.6818,43.7106)
    .. controls (309.6818,39.1466) and (324.1812,35.4468) .. (342.0675,35.4468)
    .. controls (359.9539,35.4468) and (374.4532,39.1466) .. (374.4532,43.7106)
    -- (374.4536,121.4368)
    .. controls (374.4536,126.0008) and (359.9543,129.7006) .. (342.0679,129.7006)
    .. controls (324.1816,129.7006) and (309.6822,126.0008) .. (309.6822,121.4368)
    -- (309.6818,121.4368);
\path[draw=scalarTeal,line width=0.78499bp,line cap=round] (309.6818,121.4368)
    -- (309.6818,43.7106);
\path[draw=scalarTeal,line width=0.78499bp,line cap=round] (374.4536,121.4368)
    -- (374.4536,43.7106);
\path[draw=scalarTeal,line width=0.78499bp,line cap=round,fill=scalarPanel] (342.0679,43.7106) ellipse [x radius=32.3857bp,y radius=8.2638bp];
\path[draw=scalarTeal,line width=0.78499bp,line cap=round] (309.6818,121.4368)
    .. controls (309.6818,126.0008) and (324.1812,129.7006) .. (342.0675,129.7006)
    .. controls (359.9539,129.7006) and (374.4532,126.0008) .. (374.4532,121.4368);
\path[draw=figureTint6,line width=0.62798bp,line cap=round,dash pattern=on 2.354959bp off 2.354959bp] (309.6818,121.4368)
    .. controls (309.6818,116.8727) and (324.1812,113.1729) .. (342.0675,113.1729)
    .. controls (359.9539,113.1729) and (374.4532,116.8727) .. (374.4532,121.4368);
\path[draw=figureTint7,line width=0.58874bp,line cap=round,dash pattern=on 1.56997bp off 1.56997bp] (309.6818,65.5992)
    .. controls (309.6818,61.6518) and (324.1812,58.4518) .. (342.0675,58.4518)
    .. controls (359.9539,58.4518) and (374.4532,61.6518) .. (374.4532,65.5992);
\path[draw=melonBlue,line width=0.86348bp,line cap=round] (309.6818,65.5992)
    .. controls (309.6818,69.5466) and (324.1812,72.7466) .. (342.0675,72.7466)
    .. controls (359.9539,72.7466) and (374.4532,69.5466) .. (374.4532,65.5992);
\path[draw=figureTint8,line width=0.58874bp,line cap=round,dash pattern=on 1.56997bp off 1.56997bp] (309.6818,99.9950)
    .. controls (309.6818,96.0476) and (324.1812,92.8476) .. (342.0675,92.8476)
    .. controls (359.9539,92.8476) and (374.4532,96.0476) .. (374.4532,99.9950);
\path[draw=mapPurple,line width=0.86348bp,line cap=round] (309.6818,99.9950)
    .. controls (309.6818,103.9424) and (324.1812,107.1423) .. (342.0675,107.1423)
    .. controls (359.9539,107.1423) and (374.4532,103.9424) .. (374.4532,99.9950);
\path[draw=white,line width=0.47099bp,line cap=round,fill=melonBlue] (359.9356,71.4062) ellipse [x radius=2.1095bp,y radius=2.1095bp];
\path[draw=white,line width=0.47099bp,line cap=round,fill=mapPurple] (324.8698,106.2488) ellipse [x radius=2.1095bp,y radius=2.1095bp];

\path[draw=scalarTeal,line width=0.78499bp,line cap=round] (397.4586,118.5331)
    -- (397.4586,49.5720);
\path[draw=scalarTeal,line width=0.78499bp,line cap=round,fill=scalarTeal] (397.4586,47.1158)
    .. controls (397.3814,47.4174) and (396.9435,48.7433) .. (396.4669,49.5720)
    -- (398.4504,49.5720)
    .. controls (397.9738,48.7433) and (397.5359,47.4174) .. (397.4586,47.1158);

\node[anchor=base west,text=white,scale=0.7879168,font={\FigureSans\bfseries\fontsize{10.95}{13.140}\selectfont}] at (22.64042,15.21002) {1};
\node[anchor=base west,text=melonBlue,scale=0.7879168,font={\FigureSans\bfseries\fontsize{10.95}{13.140}\selectfont}] at (36.62757,14.70734) {Bilocal pair};
\node[anchor=base west,text=white,scale=0.7879168,font={\FigureSans\bfseries\fontsize{10.95}{13.140}\selectfont}] at (158.88281,15.21002) {2};
\node[anchor=base west,text=melonBlue,scale=0.7879168,font={\FigureSans\bfseries\fontsize{10.95}{13.140}\selectfont}] at (172.86996,14.70734) {Relative space};
\node[anchor=base west,text=white,scale=0.7879168,font={\FigureSans\bfseries\fontsize{10.95}{13.140}\selectfont}] at (302.94138,15.16119) {3};
\node[anchor=base west,text=scalarTeal,scale=0.7879168,font={\FigureSans\bfseries\fontsize{10.95}{13.140}\selectfont}] at (316.92853,14.64905) {Relative cylinder};
\node[anchor=base west,text=figureNavy,scale=0.7879168,font={\normalfont\mdseries\fontsize{12}{14.400}\selectfont}] at (13.83305,46.89386) {$x_1$};
\node[anchor=base west,text=figureNavy,scale=0.7879168,font={\normalfont\mdseries\fontsize{12}{14.400}\selectfont}] at (98.60300,123.02582) {$x_2$};
\node[anchor=base west,text=melonBlue,scale=0.7879168,font={\normalfont\mdseries\fontsize{10.95}{13.140}\selectfont}] at (68.13492,66.36023) {$x=x_{12}$};
\node[anchor=base west,text=figureSlate,scale=0.7879168,font={\FigureSans\mdseries\fontsize{10}{12.000}\selectfont}] at (14.01427,149.13910) {zero total momentum: $P_{12}=0$};
\node[anchor=base west,text=figureNavy,scale=0.7879168,font={\normalfont\mdseries\fontsize{10.95}{13.140}\selectfont}] at (187.24716,113.29739) {$x=r\mathbf n$};
\node[anchor=base west,text=melonBlue,scale=0.7879168,font={\normalfont\mdseries\fontsize{10.95}{13.140}\selectfont}] at (231.68109,55.61456) {$r$};
\node[anchor=base west,text=mapPurple,scale=0.7879168,font={\normalfont\mdseries\fontsize{10.95}{13.140}\selectfont}] at (172.73364,74.04242) {$r'$};
\node[anchor=base west,text=figureSlate,scale=0.7879168,font={\normalfont\mdseries\fontsize{10}{12.000}\selectfont}] at (170.37778,137.82379) {$r>0,\qquad\mathbf n\in S^{d-1}$};
\node[anchor=base west,text=figureSlate,scale=0.7879168,font={\FigureSans\mdseries\fontsize{10}{12.000}\selectfont}] at (162.10461,148.81369) {pair exchange: $\mathbf n\mapsto-\mathbf n$};
\node[anchor=base west,text=mapPurple,scale=0.7879168,font={\normalfont\mdseries\fontsize{10}{12.000}\selectfont}] at (257.82904,61.61688) {$\tau=\log r$};
\node[anchor=base west,text=melonBlue,scale=0.7879168,font={\normalfont\mdseries\fontsize{10.95}{13.140}\selectfont}] at (381.47571,67.22852) {$\tau$};
\node[anchor=base west,text=mapPurple,scale=0.7879168,font={\normalfont\mdseries\fontsize{10.95}{13.140}\selectfont}] at (381.47571,103.30103) {$\tau'$};
\node[anchor=base west,text=scalarTeal,scale=0.7879168,font={\FigureSans\mdseries\fontsize{10}{12.000}\selectfont},rotate=90] at (407.65039,108.77948) {logarithmic time};
\node[anchor=base west,text=figureSlate,scale=0.7879168,font={\FigureSans\mdseries\fontsize{10}{12.000}\selectfont}] at (313.70828,137.55511) {Coordinates: $(\tau,\mathbf n)$};
\node[anchor=base west,text=figureSlate,scale=0.7879168,font={\FigureSans\mdseries\fontsize{10}{12.000}\selectfont}] at (298.05389,148.54501) {with $\tau=\log r$ and $\mathbf n\in S^{d-1}$};
\coordinate (bilocal-one) at (27.14395,52.64460);
\coordinate (bilocal-two) at (94.81920,115.18300);
\coordinate (bilocal-centre) at ($(bilocal-one)!0.5!(bilocal-two)$);
\fill[black] (bilocal-centre) circle[radius=1.9bp];
\node[anchor=base west,text=black,scale=0.7879168,font={\normalfont\mdseries\fontsize{12}{14.400}\selectfont}] at (49.90000,94.30000) {$X$};
\end{tikzpicture}%
\endgroup%

%% file: sections/03_robin_ladders.tex
\section{Melonic ladders as Robin propagators}
\label{sec:robin}
\subsection{Reduced dynamics select a Neumann seed}
From this section onwards, we specialise to melonic large-$N$ CFTs and their ladder sum~\cite{MS16,Witten16,KT16,GKT17}. Let $G_{ij}\equiv G(x_{ij})$ be the dressed elementary propagator and $C_0$ the leading covariance of the singlet bilocal, which we call the seed. For bosonic pairs, $C_0=G_{13}G_{24}+G_{14}G_{23}$; the fermionic Wick signs are specified in appendix~\ref{app:syk}. The leading connected two-point function of the singlet bilocal fluctuations is
\begin{equation}
\begin{split}
F_g&\equiv(1-gK)^{-1}C_0=\sum_{n=0}^\infty(gK)^nC_0,\qquad K\equiv C_0\mathcal V,\\
\mathcal V(y)&\equiv v_0|y|^{-2\nu},\qquad \nu\equiv d-2\Delta_\phi=d/2-a,
\end{split}
\label{eq:Fg}
\end{equation}
where $K$ is the rung kernel and $g$ is the ladder coupling. 
Henceforth, we shall refer to $F_g$ as the ladder. 
Here $y\equiv x_{34}$, and $v_0$ includes the model's combinatorial and index-contraction factors with the full ordered-pair measure. The zero-centre-momentum reduction is
\begin{equation}
F^{(0)}_g(x,y)\equiv \int\mathrm{d}^dY\,F_g\big(x_1,x_2;x_3,x_4\big)=\int\mathrm{d}^dY\,F_g\big(\tfrac{x}{2},-\tfrac{x}{2};Y+\tfrac{y}{2},Y-\tfrac{y}{2}\big),
\label{eq:F0}
\end{equation}
where $Y\equiv(-x_1-x_2+x_3+x_4)/2$ is the separation of the centres and the second equality follows from translational invariance. 
Translation invariance also makes composition fibrewise, $(KF)^{(0)}=K^{(0)}F^{(0)}$, on a common convergent or regulated domain. This is not a bounded projection on arbitrary bilocal $L^2$ data; appendices~\ref{app:A} and~\ref{app:syk} specify the convergence and continuation prescriptions.

The rail integral $\int\mathrm{d}^dz\,G(z)G(z-t)\propto|t|^{-2a}$ gives $C^{(0)}_0\propto[|x-y|^{-2a}+\sigma|x+y|^{-2a}]$, where $\sigma=\pm1$ is the retained pair parity. On functions $f(-y)=\sigma f(y)$, centre independence of $\mathcal V$ gives
\begin{equation}
K^{(0)}(x,y)=\frac{c_\partial\Gamma(a)}{2\pi^{d/2}\Gamma(1-\nu)}|x-y|^{-2a}|y|^{-2\nu},
\label{eq:K0}
\end{equation}
where $c_\partial$ is fixed by the seed normalisation in appendix~\ref{app:A}. The bosonic centre integral is absolutely convergent for $d/4<\Re\Delta_\phi<d/2$, and is otherwise defined by continuation from this strip. In SYK with $q\ge4$, $\Delta=1/q\le1/4$: the individual exchange integrals are infrared divergent, but their antisymmetric combination converges. We therefore combine the Wick-exchange terms before integration for all $q\ge4$. The further $a\to0$ limit applies only at $q=4$ and gives a finite logarithmic seed (appendix~\ref{app:syk}).

Using the same $W$ as in section~\ref{sec:cylinder}, with $x=e^\tau\mathbf n$ and $y=e^{\tau'}\mathbf n'$, gives the stationary cylinder kernel
\begin{equation}
\begin{split}
WK^{(0)}W^{-1}&=c_\partial G_N,\qquad G_N\equiv\frac{\Gamma(a)}{2\pi^{d/2}\Gamma(1-\nu)}P_\sigma\mathcal D_a,\\
\mathcal D_a(\tau-\tau';\mathbf n,\mathbf n')&\equiv[2\cosh(\tau-\tau')-2\mathbf n\cdot\mathbf n']^{-a},
\end{split}
\label{eq:Da}
\end{equation}
where $P_\sigma f(\mathbf n)\equiv[f(\mathbf n)+\sigma f(-\mathbf n)]/2$ and $|x-y|^2=e^{\tau+\tau'}[2\cosh(\tau-\tau')-2\mathbf n\cdot\mathbf n']$. 
The cylinder measure is $\mathrm d\tau\,\mathrm d\Omega_{d-1}$. 
As shown below and in appendix~\ref{app:B}, $G_N$ is the Neumann boundary covariance of a scalar in global Euclidean AdS$_{d+1}$, within the parameter range stated in section~\ref{sec:discussion}, with mass
\begin{equation}
m^2=\nu^2-d^2/4.
\label{eq:mass}
\end{equation}
Using $K^{(0)}=C_0^{(0)}\mathcal V$, eq.~\eqref{eq:Da} equivalently gives the zero-rung identification
\begin{equation}
v_0(|x||y|)^aC_0^{(0)}(x,y)=c_\partial G_N(\tau-\tau',\mathbf n,\mathbf n').
\label{eq:zerorung}
\end{equation}
The two external weights are the kernel form of the same conjugation by $W$: the Jacobian $\mathrm d^dy=|y|^d\mathrm d\tau'\,\mathrm d\Omega'_{d-1}$ combines with $\mathcal V(y)$ and the input factor $|y|^{-a}$ to give $v_0|y|^a$, since $d-a-2\nu=a$. 

\subsection{Robin scalar in \texorpdfstring{AdS$_{d+1}$}{AdS(d+1)} and the melonic ladder}
For a bulk scalar $\Phi$ with $0<\nu<1$, define $\Delta_\pm\equiv d/2\pm\nu$ and asymptotic coefficients by $\Phi\sim A\zeta^{\Delta_-}+B\zeta^{\Delta_+}$, with $\zeta\equiv\cos\theta$ and $\theta$ the global radial angle. At zero source, Dirichlet means $A=0$, Neumann means $B=0$, and Robin means $B=\kappa A$. In the minimal subtraction scheme, the Euclidean action is
\begin{equation}
\begin{split}
S[\Phi;j]&\equiv S_{\rm bulk}+S_{\rm ct}+S_R,\\
S_{\rm bulk}&\equiv\frac12\int_{\mathcal M}\mathrm d^{d+1}x\sqrt g\,\big[(\nabla\Phi)^2+m^2\Phi^2\big],\\
S_R&\equiv\int_{\partial\mathcal M}\mathrm d\tau\,\mathrm d\Omega_{d-1}\,\big(\nu\kappa A^2-jA\big).
\end{split}
\label{eq:action}
\end{equation}
Here $\mathcal M$ is Euclidean AdS$_{d+1}$, $S_{\rm ct}$ contains the cutoff-boundary counterterm (appendix~\ref{app:B}), and $j$ is a boundary source. After removing the cutoff, the on-shell variation is $\delta(S_{\rm bulk}+S_{\rm ct})=-2\nu\int_{\partial\mathcal M}\mathrm d\tau\,\mathrm d\Omega_{d-1}\,B\delta A$. Including $S_R$, stationarity at fixed $j$ gives $j=2\nu(\kappa A-B)$.

At fixed $(\omega,J)$, regularity at the centre for $d>1$, or fixed parity for $d=1$, determines the radial solution up to normalisation. Its boundary coefficients define the meromorphic response $\mathcal R_J(\omega)\equiv B/A$, with poles at Dirichlet frequencies. The Robin boundary-to-boundary propagator, or boundary covariance, $G_R\equiv\langle AA\rangle$, is therefore
\begin{equation}
\widehat G_R(\omega,J;\kappa)=\frac{1}{2\nu[\kappa-\mathcal R_J(\omega)]}.
\label{eq:G_R}
\end{equation}
The hat denotes the Fourier--harmonic representation, continued from Euclidean frequency $p$ by $\omega=ip$. When only one cylinder time is displayed, it denotes the separation. Setting $\kappa=0$ gives $G_N$, so eq.~\eqref{eq:G_R} implies the operator identity
\begin{equation}
G_R(\kappa)=(G_N^{-1}+2\nu\kappa)^{-1}.
\label{eq:Gkappa}
\end{equation}
The quadratic $A^2$ term implements Robin (mixed) boundary conditions~\cite{Witten01} and gives the standard double-trace resolvent~\eqref{eq:Gkappa}~\cite{GK02,HR06}; explicit boundary-chain expansions appear in global AdS$_4$~\cite{MTW20} and in the AdS/BTZ construction of Ref.~\cite{WY26}. 
Equivalently, one may impose $B = \kappa A$ directly as a self-adjoint boundary condition; the canonical two-point function of $A$ reproduces eq.~\eqref{eq:G_R}.

To identify the reduced ladder sum, multiply it on the right by $\mathcal V$: $F_g^{(0)}\mathcal V=(1-gK^{(0)})^{-1}K^{(0)}$. With $\kappa=-gc_\partial/(2\nu)$, eqs.~\eqref{eq:Da} and~\eqref{eq:Gkappa} give
\begin{equation}
\begin{aligned}
W\bigl(F_g^{(0)}\mathcal V\bigr)W^{-1}
&=(1-gc_\partial G_N)^{-1}c_\partial G_N\\
&=c_\partial(G_N^{-1}-gc_\partial)^{-1}
=c_\partial G_R(\kappa).
\end{aligned}
\label{eq:resum}
\end{equation}
As in eq.~\eqref{eq:zerorung}, the cylinder kernel on the left is $v_0(|x||y|)^aF_g^{(0)}(x,y)$. No commutation of $C_0^{(0)}$ and $\mathcal V$ is assumed. The series is first defined where it converges and then by resolvent continuation (appendix~\ref{app:A}). Thus
\begin{equation}
\boxed{v_0(|x||y|)^aF_g^{(0)}(x,y)=c_\partial G_R(\tau-\tau',\mathbf n,\mathbf n';\kappa).}
\label{eq:main}
\end{equation}
Equivalently, if $k(h,J)$ is the rung kernel eigenvalue,
\begin{equation}
\frac{k(h,J)}{1-gk(h,J)}=c_\partial\widehat G_R(h-d/2,J;\kappa).
\label{eq:mainhat}
\end{equation}
These off-shell identities are illustrated in figure~\ref{fig:rungs}. 
The on-shell energies satisfy $\mathcal R_J(\omega)=\kappa$. 
Here we consider a single ladder $F_g$. 
For melonic theories with richer constituent field content and multiple ladder sums, such as complex or SUSY SYK~\cite{Bulycheva:2017uqj,Peng:2017spg}, we expect the bulk description to extend beyond the single scalar considered here.

\input{figures/correspondence}

\subsection{Spectrum and OPE coefficients}
\label{sec:ope}
We now ask what conformal data are retained by eq.~\eqref{eq:main}. Write each family contribution as $c_m^2\mathcal B_{h_m,J_m}$, where $c_m$ is its OPE coefficient and $\mathcal B_{h,J}$ includes the external-coordinate factors and fixed singlet normalisation of $F_g$. The family functions use the same continuation across pair orderings as the ladder. Whenever the family sum can be reduced term by term under the prescription in appendix~\ref{app:A}, linearity gives
\begin{equation}
\begin{aligned}
\boxed{F_g=\sum_m c_m^2\mathcal B_{h_m,J_m},\qquad
 G_R=\sum_m c_m^2\,\widetilde{\mathcal B}_{h_m,J_m},}
\end{aligned}
\label{eq:OPEreduction}
\end{equation}
where $\widetilde{\mathcal B}_{h,J}\equiv(v_0/c_\partial)(|x||y|)^a\mathcal B_{h,J}^{(0)}$ and the superscript $(0)$ denotes the integral in eq.~\eqref{eq:F0}. 
That is, the same coefficient $c_m^2$ multiplies each reduced family. Once the block convention is fixed, the integral fixes its reduced normalisation. If the reduced families are independent, these coefficients are uniquely determined. Appendix~\ref{app:independence} gives sufficient conditions, and appendix~\ref{app:sykope} verifies them for the ordinary SYK families. For AR, appendix~\ref{app:arope} instead inverts the centre integral of the full four-point function, recovering its OPE coefficients without interchanging the OPE sum and the integral. In both examples, the scalar pole frequencies give $h_m=\omega_m+d/2$ through eq.~\eqref{eq:kR}. The divergent SYK $h=2$ mode is treated separately in appendix~\ref{app:csyk}.

%% file: figures/correspondence.tex
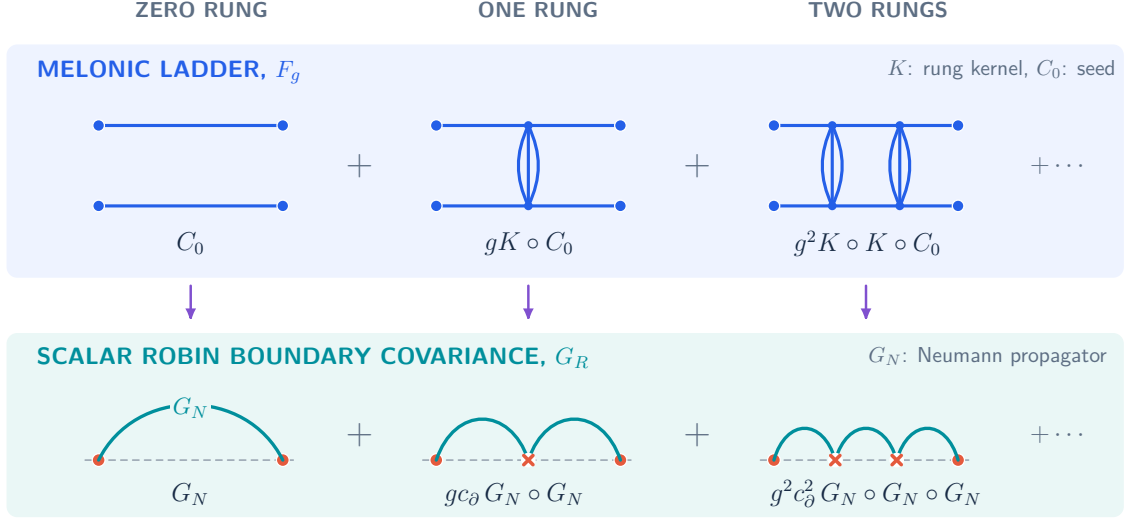
\begin{figure}[tbp]
\centering
\resizebox{\textwidth}{!}{\input{figures/artwork/correspondence.tex}}
\caption{The same boundary vertex reproduces every rung. $\circ$ denotes kernel composition with the appropriate bilocal or cylinder measure. Upper row: dressed melonic rails, with three internal lines per illustrative rung (the multiplicity is model-dependent); the fixed exchange sector is understood. Lower row: Neumann boundary covariances and integrated quadratic vertices. Each arrow implements the zero-momentum reduction and the kernel identification~\eqref{eq:main}, including its fixed normalisation.}
\label{fig:rungs}
\end{figure}

%% file: figures/artwork/correspondence.tex
\begingroup%
\input{figures/fonts/figure_fonts.tex}%
\definecolor{melonBlue}{HTML}{245FE8}%
\definecolor{figureNavy}{HTML}{182B45}%
\definecolor{figureSlate}{HTML}{5D6E83}%
\definecolor{melonPanel}{HTML}{EEF3FF}%
\definecolor{scalarTeal}{HTML}{008F9C}%
\definecolor{boundaryOrange}{HTML}{E45A3D}%
\definecolor{mapPurple}{HTML}{8050CF}%
\definecolor{scalarPanel}{HTML}{EAF7F6}%
\definecolor{footerPanel}{HTML}{F7F9FC}%
\definecolor{figureTint1}{HTML}{96A1AE}%
\begin{tikzpicture}[x=1bp,y=-1bp,inner sep=0pt,outer sep=0pt]%
\path[use as bounding box] (0,0) rectangle (416,193.3856);

\path[fill=melonPanel] (4.8090,102.3539)
    -- (4.8090,25.3309)
    .. controls (4.8090,23.1632) and (6.5662,21.4060) .. (8.7339,21.4060)
    -- (407.3813,21.4060)
    .. controls (409.5490,21.4060) and (411.3063,23.1632) .. (411.3063,25.3309)
    -- (411.3063,102.3539)
    .. controls (411.3063,104.5216) and (409.5490,106.2788) .. (407.3813,106.2788)
    -- (8.7339,106.2788)
    .. controls (6.5662,106.2788) and (4.8090,104.5216) .. (4.8090,102.3539);
\path[fill=scalarPanel] (4.8090,189.4606)
    -- (4.8090,130.3054)
    .. controls (4.8090,128.1377) and (6.5662,126.3805) .. (8.7339,126.3805)
    -- (407.3813,126.3805)
    .. controls (409.5490,126.3805) and (411.3063,128.1377) .. (411.3063,130.3054)
    -- (411.3063,189.4606)
    .. controls (411.3063,191.6283) and (409.5490,193.3856) .. (407.3813,193.3856)
    -- (8.7339,193.3856)
    .. controls (6.5662,193.3856) and (4.8090,191.6283) .. (4.8090,189.4606);

\path[draw=melonBlue,line width=1.13823bp,line cap=round] (38.3115,80.1469)
    -- (105.3166,80.1469);
\path[draw=melonBlue,line width=1.13823bp,line cap=round] (38.3115,50.8880)
    -- (105.3166,50.8880);
\path[draw=white,line width=0.47099bp,line cap=round,fill=melonBlue] (38.3115,80.1469) ellipse [x radius=2.1095bp,y radius=2.1095bp];
\path[draw=white,line width=0.47099bp,line cap=round,fill=melonBlue] (38.3115,50.8880) ellipse [x radius=2.1095bp,y radius=2.1095bp];
\path[draw=white,line width=0.47099bp,line cap=round,fill=melonBlue] (105.3166,80.1469) ellipse [x radius=2.1095bp,y radius=2.1095bp];
\path[draw=white,line width=0.47099bp,line cap=round,fill=melonBlue] (105.3166,50.8880) ellipse [x radius=2.1095bp,y radius=2.1095bp];
\path[draw=mapPurple,line width=0.90272bp,line cap=round] (71.8140,109.1824)
    -- (71.8140,117.8979);
\path[draw=mapPurple,line width=0.90272bp,line cap=round,fill=mapPurple] (71.8140,120.7955)
    .. controls (71.9081,120.4396) and (72.4411,118.8755) .. (73.0211,117.8979)
    -- (70.6070,117.8979)
    .. controls (71.1870,118.8755) and (71.7200,120.4396) .. (71.8140,120.7955);
\path[draw=figureTint1,line width=0.54949bp,line cap=round,dash pattern=on 2.354959bp off 1.9624645bp] (33.3978,172.6138)
    -- (110.2303,172.6138);
\path[draw=white,line width=0.47099bp,line cap=round,fill=boundaryOrange] (38.3115,172.6138) ellipse [x radius=2.3314bp,y radius=2.3314bp];
\path[draw=white,line width=0.47099bp,line cap=round,fill=boundaryOrange] (105.3166,172.6138) ellipse [x radius=2.3314bp,y radius=2.3314bp];
\path[draw=scalarTeal,line width=1.21672bp,line cap=round] (38.3115,172.6138)
    .. controls (51.7127,145.8119) and (91.9154,145.8119) .. (105.3166,172.6138);
\path[fill=scalarPanel] (64.4712,149.2422) rectangle (79.1568,158.0164);

\path[draw=melonBlue,line width=1.13823bp,line cap=round] (161.1542,80.1469)
    -- (228.1592,80.1469);
\path[draw=melonBlue,line width=1.13823bp,line cap=round] (161.1542,50.8880)
    -- (228.1592,50.8880);
\path[draw=white,line width=0.47099bp,line cap=round,fill=melonBlue] (161.1542,80.1469) ellipse [x radius=2.1095bp,y radius=2.1095bp];
\path[draw=white,line width=0.47099bp,line cap=round,fill=melonBlue] (161.1542,50.8880) ellipse [x radius=2.1095bp,y radius=2.1095bp];
\path[draw=white,line width=0.47099bp,line cap=round,fill=melonBlue] (228.1592,80.1469) ellipse [x radius=2.1095bp,y radius=2.1095bp];
\path[draw=white,line width=0.47099bp,line cap=round,fill=melonBlue] (228.1592,50.8880) ellipse [x radius=2.1095bp,y radius=2.1095bp];
\path[fill=melonBlue] (194.6567,80.1469) ellipse [x radius=1.5700bp,y radius=1.5700bp];
\path[fill=melonBlue] (194.6567,50.8880) ellipse [x radius=1.5700bp,y radius=1.5700bp];
\path[draw=melonBlue,line width=1.13823bp,line cap=round] (194.6567,80.1469)
    .. controls (189.8345,69.8058) and (189.8345,61.2291) .. (194.6567,50.8880);
\path[draw=melonBlue,line width=1.13823bp,line cap=round] (194.6567,80.1469)
    -- (194.6567,50.8880);
\path[draw=melonBlue,line width=1.13823bp,line cap=round] (194.6567,80.1469)
    .. controls (199.4789,69.8058) and (199.4789,61.2291) .. (194.6567,50.8880);
\path[draw=mapPurple,line width=0.90272bp,line cap=round] (194.6567,109.1824)
    -- (194.6567,117.8979);
\path[draw=mapPurple,line width=0.90272bp,line cap=round,fill=mapPurple] (194.6567,120.7955)
    .. controls (194.7508,120.4396) and (195.2838,118.8755) .. (195.8638,117.8979)
    -- (193.4496,117.8979)
    .. controls (194.0296,118.8755) and (194.5627,120.4396) .. (194.6567,120.7955);
\path[draw=figureTint1,line width=0.54949bp,line cap=round,dash pattern=on 2.354959bp off 1.9624645bp] (156.2404,172.6138)
    -- (233.0729,172.6138);
\path[draw=white,line width=0.47099bp,line cap=round,fill=boundaryOrange] (161.1542,172.6138) ellipse [x radius=2.3314bp,y radius=2.3314bp];
\path[draw=white,line width=0.47099bp,line cap=round,fill=boundaryOrange] (228.1592,172.6138) ellipse [x radius=2.3314bp,y radius=2.3314bp];
\path[draw=scalarTeal,line width=1.21672bp,line cap=round] (161.1542,172.6138)
    .. controls (167.8547,152.7357) and (187.9561,152.7357) .. (194.6567,172.6138)
    .. controls (201.3573,152.7357) and (221.4587,152.7357) .. (228.1592,172.6138);
\path[fill=white] (194.6567,172.6138) ellipse [x radius=2.6689bp,y radius=2.6690bp];
\path[draw=boundaryOrange,line width=1.29521bp,line cap=round] (192.9817,174.2889)
    -- (196.3318,170.9388);
\path[draw=boundaryOrange,line width=1.29521bp,line cap=round] (192.9817,170.9388)
    -- (196.3318,174.2889);

\path[draw=melonBlue,line width=1.13823bp,line cap=round] (283.9968,80.1469)
    -- (351.0019,80.1469);
\path[draw=melonBlue,line width=1.13823bp,line cap=round] (283.9968,50.8880)
    -- (351.0019,50.8880);
\path[draw=white,line width=0.47099bp,line cap=round,fill=melonBlue] (283.9968,80.1469) ellipse [x radius=2.1095bp,y radius=2.1095bp];
\path[draw=white,line width=0.47099bp,line cap=round,fill=melonBlue] (283.9968,50.8880) ellipse [x radius=2.1095bp,y radius=2.1095bp];
\path[draw=white,line width=0.47099bp,line cap=round,fill=melonBlue] (351.0019,80.1469) ellipse [x radius=2.1095bp,y radius=2.1095bp];
\path[draw=white,line width=0.47099bp,line cap=round,fill=melonBlue] (351.0019,50.8880) ellipse [x radius=2.1095bp,y radius=2.1095bp];
\path[fill=melonBlue] (305.2150,80.1469) ellipse [x radius=1.5700bp,y radius=1.5700bp];
\path[fill=melonBlue] (305.2150,50.8880) ellipse [x radius=1.5700bp,y radius=1.5700bp];
\path[draw=melonBlue,line width=1.13823bp,line cap=round] (305.2150,80.1469)
    .. controls (300.3929,69.8058) and (300.3929,61.2291) .. (305.2150,50.8880);
\path[draw=melonBlue,line width=1.13823bp,line cap=round] (305.2150,80.1469)
    -- (305.2150,50.8880);
\path[draw=melonBlue,line width=1.13823bp,line cap=round] (305.2150,80.1469)
    .. controls (310.0372,69.8058) and (310.0372,61.2291) .. (305.2150,50.8880);
\path[fill=melonBlue] (329.7837,80.1469) ellipse [x radius=1.5700bp,y radius=1.5700bp];
\path[fill=melonBlue] (329.7837,50.8880) ellipse [x radius=1.5700bp,y radius=1.5700bp];
\path[draw=melonBlue,line width=1.13823bp,line cap=round] (329.7837,80.1469)
    .. controls (324.9615,69.8058) and (324.9615,61.2291) .. (329.7837,50.8880);
\path[draw=melonBlue,line width=1.13823bp,line cap=round] (329.7837,80.1469)
    -- (329.7837,50.8880);
\path[draw=melonBlue,line width=1.13823bp,line cap=round] (329.7837,80.1469)
    .. controls (334.6058,69.8058) and (334.6058,61.2291) .. (329.7837,50.8880);
\path[draw=mapPurple,line width=0.90272bp,line cap=round] (317.4994,109.1824)
    -- (317.4994,117.8979);
\path[draw=mapPurple,line width=0.90272bp,line cap=round,fill=mapPurple] (317.4994,120.7955)
    .. controls (317.5934,120.4396) and (318.1264,118.8755) .. (318.7065,117.8979)
    -- (316.2923,117.8979)
    .. controls (316.8723,118.8755) and (317.4053,120.4396) .. (317.4994,120.7955);
\path[draw=figureTint1,line width=0.54949bp,line cap=round,dash pattern=on 2.354959bp off 1.9624645bp] (279.0831,172.6138)
    -- (355.9156,172.6138);
\path[draw=white,line width=0.47099bp,line cap=round,fill=boundaryOrange] (283.9968,172.6138) ellipse [x radius=2.3314bp,y radius=2.3314bp];
\path[draw=white,line width=0.47099bp,line cap=round,fill=boundaryOrange] (351.0019,172.6138) ellipse [x radius=2.3314bp,y radius=2.3314bp];
\path[draw=scalarTeal,line width=1.21672bp,line cap=round] (283.9968,172.6138)
    .. controls (288.4637,157.2026) and (301.8649,157.2026) .. (306.3318,172.6138)
    .. controls (310.7988,157.2026) and (324.1999,157.2026) .. (328.6669,172.6138)
    .. controls (333.1338,157.2026) and (346.5349,157.2026) .. (351.0019,172.6138);
\path[fill=white] (306.3318,172.6138) ellipse [x radius=2.6689bp,y radius=2.6690bp];
\path[draw=boundaryOrange,line width=1.29521bp,line cap=round] (304.6568,174.2889)
    -- (308.0069,170.9388);
\path[draw=boundaryOrange,line width=1.29521bp,line cap=round] (304.6568,170.9388)
    -- (308.0069,174.2889);
\path[fill=white] (328.6669,172.6138) ellipse [x radius=2.6689bp,y radius=2.6690bp];
\path[draw=boundaryOrange,line width=1.29521bp,line cap=round] (326.9918,174.2889)
    -- (330.3419,170.9388);
\path[draw=boundaryOrange,line width=1.29521bp,line cap=round] (326.9918,170.9388)
    -- (330.3419,174.2889);


\node[anchor=base west,text=melonBlue,scale=0.7879168,font={\FigureSans\bfseries\fontsize{10.95}{13.140}\selectfont}] at (15.85642,33.20258) {MELONIC LADDER, $F_g$};
\node[anchor=base west,text=figureSlate,scale=0.7879168,font={\FigureSans\mdseries\fontsize{10}{12.000}\selectfont}] at (325.29907,32.34613) {$K$: rung kernel, $C_0$: seed};
\node[anchor=base west,text=scalarTeal,scale=0.7879168,font={\FigureSans\bfseries\fontsize{10.95}{13.140}\selectfont}] at (15.85642,137.72882) {SCALAR ROBIN BOUNDARY COVARIANCE, $G_R$};
\node[anchor=base west,text=figureSlate,scale=0.7879168,font={\FigureSans\mdseries\fontsize{10}{12.000}\selectfont}] at (318.39047,136.80222) {$G_N$: Neumann propagator};
\node[anchor=base west,text=figureSlate,scale=0.7879168,font={\FigureSans\bfseries\fontsize{10}{12.000}\selectfont}] at (51.53344,11.51270) {ZERO RUNG};
\node[anchor=base west,text=figureSlate,scale=0.7879168,font={\FigureSans\bfseries\fontsize{10}{12.000}\selectfont}] at (176.29349,11.51270) {ONE RUNG};
\node[anchor=base west,text=figureSlate,scale=0.7879168,font={\FigureSans\bfseries\fontsize{10}{12.000}\selectfont}] at (296.34018,11.51270) {TWO RUNGS};
\node[anchor=base west,text=figureNavy,scale=0.7879168,font={\normalfont\mdseries\fontsize{12}{14.400}\selectfont}] at (66.64890,96.28503) {$C_0$};
\node[anchor=base west,text=scalarTeal,scale=0.7879168,font={\normalfont\mdseries\fontsize{10.95}{13.140}\selectfont}] at (65.25507,155.90219) {$G_N$};
\node[anchor=base west,text=figureNavy,scale=0.7879168,font={\normalfont\mdseries\fontsize{12}{14.400}\selectfont}] at (64.87266,187.41110) {$ G_N$};
\node[anchor=base west,text=figureNavy,scale=0.7879168,font={\normalfont\mdseries\fontsize{12}{14.400}\selectfont}] at (178.67615,96.07544) {$gK\circ C_0$};
\node[anchor=base west,text=figureNavy,scale=0.7879168,font={\normalfont\mdseries\fontsize{12}{14.400}\selectfont}] at (164.17224,187.46310) {$gc_\partial\,G_N\circ  G_N$};
\node[anchor=base west,text=figureNavy,scale=0.7879168,font={\normalfont\mdseries\fontsize{12}{14.400}\selectfont}] at (291.21555,96.58997) {$g^2K\circ K \circ C_0$};
\node[anchor=base west,text=figureNavy,scale=0.7879168,font={\normalfont\mdseries\fontsize{12}{14.400}\selectfont}] at (282.58707,187.40640) {$g^2 c_\partial^2\,G_N\circ G_N \circ  G_N$};
\node[anchor=base west,text=figureSlate,scale=0.7879168,font={\normalfont\mdseries\fontsize{17.28}{20.736}\selectfont}] at (128.35172,69.02222) {$+$};
\node[anchor=base west,text=figureSlate,scale=0.7879168,font={\normalfont\mdseries\fontsize{17.28}{20.736}\selectfont}] at (128.35172,166.84796) {$+$};
\node[anchor=base west,text=figureSlate,scale=0.7879168,font={\normalfont\mdseries\fontsize{17.28}{20.736}\selectfont}] at (251.19235,69.02222) {$+$};
\node[anchor=base west,text=figureSlate,scale=0.7879168,font={\normalfont\mdseries\fontsize{17.28}{20.736}\selectfont}] at (251.19235,166.84796) {$+$};
\node[anchor=base west,text=figureSlate,scale=0.7879168,font={\normalfont\mdseries\fontsize{12}{14.400}\selectfont}] at (377.19809,67.98608) {$+\cdots$};
\node[anchor=base west,text=figureSlate,scale=0.7879168,font={\normalfont\mdseries\fontsize{12}{14.400}\selectfont}] at (377.19809,165.81265) {$+\cdots$};
\end{tikzpicture}%
\endgroup%

%% file: sections/05_examples.tex
\section{Examples}
\label{sec:examples}
In this section, we study the correspondence for two specific melonic CFTs. 
We pick a model with disorder, the SYK model, and a model without disorder, the $d=2$ long-range AR model. 

\subsection{SYK and an \texorpdfstring{AdS$_2$}{AdS2} scalar}
\label{sec:syk}
For $N$ Majorana fermions and even $q\ge2$, the Euclidean SYK action is
\begin{equation}
S_{\rm SYK}\equiv\int\mathrm{d}T \,\Big[\tfrac12\sum_{i=1}^N\psi_i\partial_T\psi_i+i^{\frac{q}{2}}\!\!\sum_{1\le i_1<\cdots<i_q\le N}\!j_{i_1\cdots i_q}\psi_{i_1}\cdots\psi_{i_q}\Big].
\label{eq:SYK}
\end{equation}
The $j_{i_1\cdots i_q}$ are sampled from real Gaussians with variance $\langle j_{i_1\cdots i_q}^2\rangle=J^2(q-1)!/N^{q-1}$. 
Here $J$ is not the harmonic label from earlier and $T$ is microscopic Euclidean time.
In the conformal (strong-coupling) large-$N$ regime, the SYK propagator is given by $G_c(T)=b\,{\rm sgn}(T)/|T|^{2\Delta}$, with $\Delta=1/q$ and $J^2b^q\pi=(\tfrac12-\Delta)\tan(\pi\Delta)$; see Ref.~\cite{MS16} for details. 

The SYK ladder selects an odd ($J=1$) scalar field in Euclidean AdS$_2$ with mass and Robin boundary condition given by
\begin{equation}
\nu=1-2/q,\qquad m^2=\nu^2-1/4,\qquad \kappa=4^{-\nu}(q-1).
\label{eq:SYKdata}
\end{equation}
Equation~\eqref{eq:SYKdata} corresponds to $g=1$. For $0\le g<1$, the ladder also has a Robin interpretation on the AdS$_2$ strip, with the coupling given in appendix~\ref{app:csyk}.
Let $t$ be the Lorentzian time and $x\in[-\pi/2,\pi/2]$ be the radial coordinate. 
The metric and equation of motion satisfied by the free scalar in global AdS$_2$ are
\begin{equation}
ds^2=\frac{-dt^2+dx^2}{\cos^2x}, \qquad \Big(-\partial_t^2+\partial_x^2-\frac{m^2}{\cos^2x}\Big)\Phi=0,
\label{eq:AdS2}
\end{equation}
respectively. 
For $q>2$, the field obeys $\Phi\sim A_\pm(\cos x)^{1/2-\nu}+B_\pm(\cos x)^{1/2+\nu}$ as $x\to\pm\pi/2$, with $B_\pm=\kappa A_\pm$ and $\Phi(t,-x)=-\Phi(t,x)$. In this range, $0<\nu<1$ and the positive Robin coupling in eq.~\eqref{eq:SYKdata} gives a positive spatial operator in the odd sector, so the scalar has a stable Fock vacuum (appendix~\ref{app:B}). The $q=4$ scalar is massless. At $q=2$, the mass reaches the Breitenlohner--Freedman bound and the two boundary powers coalesce; logarithmic boundary data replace the Robin formulas above.

The conformal SYK ladder sum diverges in the $h=2$ sector at $g=1$~\cite{MS16}. We keep $g$ variable during the centre reduction and take $g\to1$ only afterwards, obtaining a finite result. Appendix~\ref{app:csyk} shows that this agrees with reducing the regulated conformal ladder of Maldacena and Stanford~\cite{MS16}. Its scalar frequency is $\omega=3/2$, so it is not a scalar zero mode. Finite-coupling $h=2$ corrections are not included~\cite[section~III.C.4]{MS16}.

\subsection{AR model and an \texorpdfstring{AdS$_3$}{AdS3} scalar}
\label{sec:ar}
The AR model was introduced by Amit and Roginsky~\cite{AR79}. For its mass-tuned long-range version, we use the field and coupling notation of Benedetti and Delporte~\cite{BD20}, specialised to $d=2$:
\begin{equation}
\begin{split}
S\equiv\int\mathrm{d}^2x \, \Big[&\tfrac12\sum_m\big(\phi^m(-\partial^2)^{1/3}\phi_m+\lambda_2\phi_m\phi^m\big)\\
&+\frac{\lambda\sqrt{2j+1}}{3!}\sum_{m_1,m_2,m_3}
\begin{pmatrix}j&j&j\\m_1&m_2&m_3\end{pmatrix}
\phi^{m_1}\phi^{m_2}\phi^{m_3}\Big].
\end{split}
\label{eq:AR}
\end{equation}
The $N\equiv2j+1$ real degrees of freedom form an internal spin-$j$ representation of $SO(3)$, with $j$ even and $m=-j,\ldots,j$. The index conventions and critical mass tuning are specified in appendix~\ref{app:ar}. At the melonic large-$N$ saddle, $G(p)=\mathcal Z_{\rm LR}C(p)$, with $C(p)\equiv|p|^{-2/3}$.
The physical rung strength is $g\equiv\lambda^2\mathcal Z_{\rm LR}^3$. Note that this is not the rescaled coupling of Ref.~\cite{BD20}, which differs by a square root.

The reduced ladder selects an even scalar in Euclidean AdS$_3$ with
\begin{equation}
\nu=2/3,\qquad m^2=-5/9,\qquad \kappa=-\frac{3 g}{8\pi}.
\label{eq:ARdata}
\end{equation}
Let $t$ be the Lorentzian time, $0\le r<\pi/2$ the radial coordinate, and $\varphi\sim\varphi+2\pi$ the angular coordinate. The metric and equation of motion satisfied by the free scalar are
\begin{equation}
\begin{split}
&ds^2=\frac{-dt^2+dr^2+\sin^2r\,d\varphi^2}{\cos^2r},\\
&\Big[-\partial_t^2+\frac{1}{\tan r}\partial_r(\tan r\,\partial_r)+\frac{\partial_\varphi^2}{\sin^2r}+\frac{5}{9\cos^2r}\Big]\Phi=0,
\end{split}
\label{eq:ARmetric}
\end{equation}
respectively. The field obeys $\Phi\sim A(\cos r)^{1/3}+B(\cos r)^{5/3}$, $B=\kappa A$, and $\Phi(t,r,\varphi+\pi)=\Phi(t,r,\varphi)$.

For 
\begin{equation}
    0\le g<g_c\equiv-\frac{8\pi}{3}\mathcal R_{J=0}(\omega=0)
=-\frac{8\pi\,\Gamma(-\tfrac23)\Gamma(\tfrac56)^2}{3\Gamma(\tfrac23)\Gamma(\tfrac16)^2}
=1.02236136\ldots,
\label{eq:ARcritical}
\end{equation}
the scalar spatial operator is positive. Let $\omega_{00}$ denote the frequency of the lowest radial mode ($n=0$) in the $J=0$ sector. At $g_c$, this frequency vanishes and the dimensions $h_\pm=1\pm\omega_{00}$ collide. 
Beyond threshold, $\omega_{00}^2=-\gamma^2<0$ and $h_\pm=1\pm i\gamma$, consistent with the field-theory instability~\cite{BD20,Ben21}. The scalar mass remains above the Breitenlohner--Freedman bound: this instability comes from the Robin boundary condition. Appendix~\ref{app:B} gives the stability criterion, and appendix~\ref{app:ar} fixes the ladder normalisation.

\begin{figure}[tbp]
\centering
\includegraphics[width=\textwidth]{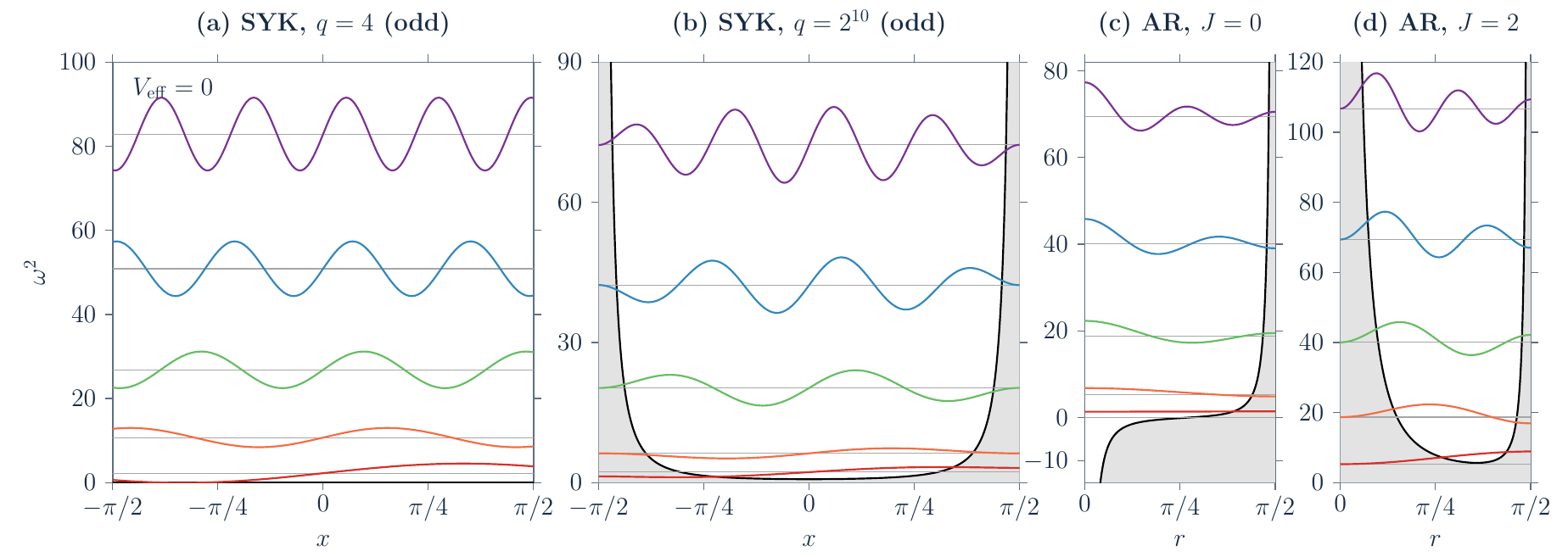}
\caption{Effective potentials and exact rescaled radial energy eigenstates for the examples discussed in the text. Black denotes the effective potential $V_{\rm eff}$, grey horizontal lines denote the levels $\omega_n^2$, and coloured curves denote the corresponding rescaled radial energy eigenstates, $\bar u$. For SYK we plot $\bar u_n(x)\equiv(\cos x)^{\nu-1/2}u_n(x)$, while for AR we plot $\bar u_{nJ}(r)\equiv(\cos r)^{-1/3}u_{nJ}(r)$. The coloured eigenstates are shifted to their respective $\omega_n^2$ levels and scaled for visibility. Panels (a) and (b) show the odd SYK sector on $-\pi/2\le x\le\pi/2$, while panels (c) and (d) show the AR model with $g=1$ on $0\le r\le\pi/2$.}
\label{fig:pot}
\end{figure}

Figure~\ref{fig:pot} displays the effective potentials and exact rescaled radial modes for both examples. 
The bulk scalars are free; the effective potentials arise from writing their radial wave equations in Schr\"odinger form.

%% file: sections/06_discussion.tex
\section{Discussion}
\label{sec:discussion}
Using radial quantisation of relative coordinates and integrating out the bilocal centres, we have related conformal melonic ladder sums to Robin boundary-to-boundary propagators of a single free bulk scalar in AdS$_{d+1}$. The cylinder is the conformal boundary; Neumann propagators joined by quadratic boundary insertions reproduce the ladder expansion. 
At fixed spacetime metric and mass, the Robin coefficient selects the allowed modes, whose bulk radial dependence follows from regularity or parity and normalisation.
We caution that radial quantisation here refers to the reduced pair Casimir, not to a full state--operator correspondence. 
Multiparticle states of the scalar have not been identified with multi-trace operators, since energies $\omega=h-d/2$ do not add like dimensions.

\paragraph{Range of parameters.}\label{par:domain}
The free-scalar interpretation established here requires $0<\nu=d-2\Delta_\phi<1$, with $a+J>0$ in each retained Neumann sector and a positive Robin spatial operator (appendix~\ref{app:B}). These conditions hold for SYK with $q>2$ in the odd sector and for the AR model with $0\leq g<g_c$; at $q=4$ ($a=0$), the odd SYK seed is given by the finite logarithmic limit in appendix~\ref{app:syk}. Outside this range, meromorphic continuation of the response does not by itself establish a stationary free-field interpretation.

\paragraph{Broken AdS generators and conformal data.}\label{sec:data}
A constant Robin coupling with $\nu\kappa\neq0$ breaks $2d$ AdS$_{d+1}$ generators: the two boundary coefficients have different conformal \mbox{weights} (appendix~\ref{app:B}). Time translations and rotations survive, giving the continuous cylinder \mbox{isometries} $\mathbb R\times SO(d)$. In SYK, the two noncompact AdS$_2$ generators also fail to preserve the odd parity sector. Nevertheless, section~\ref{sec:ope} shows that the reduced ladder retains the dimensions and OPE coefficients of the ordinary SYK families and of the stable AR ladder.  
Full auxiliary AdS symmetry is therefore not required to encode these data.

\paragraph{The DGJS construction.}
A single scalar in a three-dimensional spacetime conformal to AdS$_2\times S^1/\mathbb{Z}_2$, with a nontrivial potential, also reproduces the SYK bilinear spectrum and the bilocal propagator~\cite{DJS17,DGJS17a}. Appendix~\ref{app:dgjs} identifies its interval problem with the folded radial problem here. The map preserves the one-dimensional norms and relates their interval-centre evaluation to our boundary coefficient. The bilocal propagator comparison in Ref.~\cite{DGJS17a} uses nonstandard Poincar\'e-horizon boundary conditions in AdS$_2$ and finds an overall normalisation ratio independent of the mode, checked numerically for general $q$.

The coordinate interpretations differ. Reference~\cite{DGJS17a} retains the bilocal centre $s\equiv(T_1+T_2)/2$ and half-separation $z\equiv(T_1-T_2)/2$ as coordinates of an AdS$_2$ factor, where $T_i$ are microscopic SYK times. Its interval eigenvalues give a Kaluza--Klein tower with $M_k^2=k^2-\tfrac14=h(h-1)$. Here the centre is integrated out, $\tau\equiv\log|T_{12}|$ is cylinder time, and $k$ is the global energy of one scalar at fixed mass. The interval map therefore identifies self-adjoint spectral problems; it does not identify the spacetimes or the interacting theories.

\paragraph{Folded-string.}
This work was inspired by the folded-string construction in AdS$_2$ with imaginary radius squared~\cite{Vegh24,Vegh25}. Its momentum-fraction wavefunctions obey fractional-power Robin boundary conditions~\cite{TV26}. We show in appendix~\ref{app:foldedstring} that the folded-string momentum fraction and wavefunctions map to the radial coordinate and modes of our AdS$_2$ scalar, respectively.

\paragraph{Other ladder sums.}
Wilson-line ladder sums can also be recast as one-dimensional
Schr\"odinger problems~\cite{ESZ99,CHMS12,KKKN17}.
In these examples, the ladder coupling enters the Schr\"odinger
potential; here it enters the Robin boundary condition of a scalar
with fixed bulk mass.

\paragraph{Other constructions.}
Under the required assumptions on the spectral data and continuum limit, certain generalised free fields admit free bulk reconstructions through an HKLL-type construction or inverse scattering~\cite{NQ23,N26}. Bilocal collective fields map to the higher-spin fields of AdS$_{d+1}$~\cite{DJ03,dMKJSY18,ACU20}. The fishnet graph-building operator yields a chain of particles in AdS$_5$~\cite{GS19,GS19b}. Kinematic space keeps the pair centre and carries one field per family~\cite{CLMMS16,dBHHM16}.

Long-range models also admit local realisations as a free field coupled to a boundary or defect~\cite{PRvRZ15,GK19}; there the auxiliary space extends physical space, whereas here it is built on the relative cylinder.

It has been argued that melonic large-$N$ limits develop no extra dimension from gauge-index combinatorics~\cite{dMK19}; that analysis instead points to a bilocal description, while the auxiliary dimension here appears in the reduced bilocal fluctuation problem. Double-scaled SYK admits bulk descriptions via chord Hilbert spaces~\cite{BNS18,Lin22}.

\paragraph{Robin data and complex dimensions.}
In the standard bulk dictionary, a collision of $h$ and $d-h$ at $h=d/2$ saturates the Breitenlohner--Freedman bound, and continuation to complex dimensions violates it~\cite{PomR08,KLSS09}.  
Since $\omega^2=(h-d/2)^2=m_h^2-m_{\rm BF}^2$, with $m_h^2=h(h-d)$, the same collision appears here as a Robin zero mode at fixed scalar mass. The AR example realises this mechanism. A related boundary-condition instability occurs for a Robin scalar in global AdS$_4$~\cite{MTW20}.

A free scalar of generic mass in de Sitter space also has two independent asymptotic fall-offs at a spacelike conformal boundary, which enter the dS/CFT boundary description~\cite{Strominger:2001pn}. The Robin AdS--CFT connection developed here may therefore be useful for investigating analogous relations between boundary data and CFT correlators in de Sitter space.

\paragraph{Extensions.}
AdS$_2$ is also natural for SYK because adding Jackiw--Teitelboim (JT) gravity with a minimally coupled scalar supplies a Schwarzian gravitational boundary action under the usual nearly-AdS$_2$ boundary conditions~\cite{MSY16}. The same type of action governs the low-energy reparametrisation mode of SYK~\cite{MS16}. Here the boundary time is $\tau$; relating its reparametrisations to those of microscopic time $T$, matching the Schwarzian coefficient, and extending the Robin condition to the fluctuating boundary remain to be done.

Castro et al.~\cite{CCFZ23} proposed Wilson spools to express scalar one-loop determinants as Wilson-loop observables in three-dimensional gravity. Recent constructions apply this to JT gravity~\cite{Fliss25} and to smooth, cusp-free hyperbolic quotients in AdS$_3$ gravity~\cite{BFK25}, up to local counterterms. It would be interesting to extend these constructions to the Robin scalars studied here and compare the resulting gravitational corrections with finite-$N$ effects in the melonic models. This requires incorporating the mixed boundary condition and identifying how the gravitational expansion maps to the $1/N$ expansion.

The surviving time-translation and rotation symmetries suggest investigating Schwarz\-schild--AdS$_{d+1}$ for $d>2$~\cite{Witten98thermal} and the non-rotating BTZ black hole for $d=2$. Wang and Yang's Robin propagator~\cite{WY26} provides a starting point for the latter. A horizon introduces temperature and, with ingoing boundary conditions, quasinormal poles, so these symmetries alone do not establish a ladder correspondence. Its reduced kernel would have to be derived. 
It is natural to ask whether the single scalar can also reproduce higher-point data, including the cubic couplings of Ref.~\cite{GR17cubic}, which correspond to the subleading (in $N$) OPE coefficients of SYK bilinear operators. 
Another open question is whether the correspondence extends beyond melonic CFTs, for example to planar or vector theories.

%% file: appendices/A_reduction.tex
\section{Zero-momentum reduction and ladder resummation}
\label{app:A}

We derive the centre reduction and relate the reduced kernel eigenvalues to the scalar response.

\paragraph{Pair Casimir.}We define the centre integral of pair $(1,2)$ by $(\mathcal I f)(x)\equiv\int\mathrm{d}^dX\,f(X,x)$. For a translation-invariant four-point kernel with the second centre fixed at zero, this gives the same result as the second-pair integral in eq.~\eqref{eq:F0} after setting $X=-Y$. To perform the reduction, we rewrite eq.~\eqref{eq:C12full} as
\begin{equation}
C_{12}=D_{12}(D_{12}+d)+\mathbf L_{12}^2-P_{12}\cdot K_{12}.
\label{eq:orderedcasimir}
\end{equation}
On Schwartz functions, $P_{12}\cdot K_{12}$ is a total centre derivative, and integration by parts gives
\begin{equation}
\mathcal I D_{12}=(D_{\rm rel}-d)\mathcal I,
\qquad \mathcal I C_{12}=C_{\rm rel}\mathcal I.
\label{eq:integratedcasimir}
\end{equation}
Extension to the conformal correlators and family contributions requires the common distributional prescription described below.

\paragraph{Independence of reduced families.}\label{app:independence}
Let $\widetilde C\equiv WC_{\rm rel}W^{-1}$ and write $\widetilde{\mathcal B}_m\equiv\widetilde{\mathcal B}_{h_m,J_m}$. Suppose the completed families satisfy their homogeneous pair-Casimir equations, including across cuts and at integration endpoints, and lie in the domain of eq.~\eqref{eq:integratedcasimir}. On a common open time-ordering region, their reduced images then satisfy
\begin{equation}
\widetilde C\widetilde{\mathcal B}_m=\lambda_m\widetilde{\mathcal B}_m,
\qquad \lambda_m\equiv h_m(h_m-d)+J_m(J_m+d-2).
\label{eq:reducedfamilyindependence}
\end{equation}
Assume that the reduced families are nonzero and the $\lambda_m$ are distinct. Applying $\prod_{n\ne m}(\widetilde C-\lambda_n)$ to a finite relation $\sum_n a_n\widetilde{\mathcal B}_n=0$ gives $a_m\prod_{n\ne m}(\lambda_m-\lambda_n)\widetilde{\mathcal B}_m=0$. It follows that $a_m=0$, so the reduced families are linearly independent. This argument requires no choice of energy branch and applies away from contact terms at coincident cylinder times. 

For infinite sums, the reduction must also commute with the chosen summation prescription. If the reduced expansion has distinct positive frequencies tending to infinity and converges absolutely for every $\tau\geq\tau_0>0$, its exponential coefficients are unique: after any required harmonic projection, successive large-$\tau$ limits isolate them. Appendix~\ref{app:sykope} verifies these conditions for the completed SYK families. Alternatively, an inverse of the centre integral recovers the full correlator before its OPE is expanded. Appendix~\ref{app:arope} uses this route for AR.

\paragraph{Pair generators.}
The pair Casimir survives the reduction, but the full pair-generator action does not. For $f(x)$ independent of the centre, the special conformal action is
\begin{equation}
 K_{12,\mu}f
 =2X_\mu D_{\rm rel}f+2x_\mu X\cdot\partial_x f
 -2(X\cdot x)\partial_{x^\mu}f;
 \qquad d=1:\ K_{12}f=2X D_{\rm rel}f,
 \label{eq:Knonclosure}
\end{equation}
which generally reintroduces dependence on the centre. Hence the Fourier sector selected by zero momentum is not preserved by special conformal transformations.

\paragraph{Composition and chain integral.}For translation-invariant bilocal kernels $K_1$ and $K_2$, the reduction~\eqref{eq:F0} obeys $(K_1K_2)^{(0)}=K_1^{(0)}K_2^{(0)}$. Writing the intermediate centre as $Z$, we set $Y'\equiv Y-Z$ to separate the two centre integrals. Fubini's theorem gives the factorisation for absolutely convergent integrals. For homogeneous kernels, we establish the identity where all integrals converge and then analytically continue both sides. If regularisation is needed, we use the same prescription throughout; it must allow the centre shifts and interchange of integrals. All compositions act in the retained parity sector. We first sum the ladder where its geometric series converges and then continue the result as a resolvent.

To fix the seed normalisation, we use the standard chain integral~\cite[eq.~(2.19)]{GKT17} for $G(x)=b|x|^{-2\Delta_\phi}$:
\begin{equation}
\begin{split}
\int\mathrm{d}^dz\,G(z)G(z-t)=h_\phi|t|^{-2a}, \qquad
h_\phi\equiv\frac{b^2\pi^{d/2}\,\Gamma(\tfrac d2-\Delta_\phi)^2\,\Gamma(a)}{\Gamma(\Delta_\phi)^2\,\Gamma(\nu)}.
\end{split}
\label{eq:hphi}
\end{equation}
This integral converges for $d/4<\Re\Delta_\phi<d/2$. Outside this strip we use the same continuation in all compositions; appendix~\ref{app:syk} specifies the fermionic prescription.
The two Wick terms give $C^{(0)}_0=\epsilon h_\phi[|x-y|^{-2a}+\sigma|x+y|^{-2a}]$, where $\epsilon$ is the seed sign and $\sigma=\pm1$ the retained pair parity. On that sector, the two terms contribute equally to $K^{(0)}=C_0^{(0)}\mathcal V$. Matching eq.~\eqref{eq:K0} fixes
\begin{equation}
c_\partial=\frac{4\epsilon\pi^{d/2}v_0h_\phi\Gamma(1-\nu)}{\Gamma(a)}.
\label{eq:cBoundaryNormalisation}
\end{equation}

\paragraph{Eigenvalues.}Centre reduction preserves the eigenvalue whenever the reduced eigenfunction is nonzero. For an eigenfunction $\Psi_{h,J}$ of $K$, the relation $\mathcal I K=K^{(0)}\mathcal I$ gives
\begin{equation}
K^{(0)}(\mathcal I\Psi_{h,J})
=\mathcal I(K\Psi_{h,J})
=k(h,J)\mathcal I\Psi_{h,J},
\label{eq:kernelEigenvaluePreservation}
\end{equation}
with the common continuation described above. For $d\ge2$, we use the known conformal-kernel eigenvalue~\cite[eqs.~(2.19), (4.9)]{GKT17}, continued in $\Delta_\phi$ and written with the normalisation $c_\partial$ fixed above:
\begin{equation}
\begin{aligned}
k(h,J)&=\frac{c_\partial\Gamma(\nu)}{2\Gamma(1-\nu)}
\prod_\pm\frac{\Gamma[(a+J\pm\omega)/2]}{\Gamma[(d-a+J\pm\omega)/2]}\\
&=-\frac{c_\partial}{2\nu\mathcal R_J(\omega)},\qquad \omega=h-d/2,
\end{aligned}
\label{eq:kR}
\end{equation}
where the second equality follows from eq.~\eqref{eq:RJ}. The fermionic $d=1$ eigenvalue~\cite[eq.~(3.35)]{MS16} gives the odd-sector version of this relation, with the normalisation in appendix~\ref{app:syk}. Substituting this eigenvalue into the ladder resolvent gives eq.~\eqref{eq:mainhat}.

%% file: appendices/B_scalar.tex
\section{The Robin scalar}
\label{app:B}

\paragraph{Radial response.}The global metric is $ds^2=\sec^2\theta\,(-dt^2+d\theta^2+\sin^2\theta\,d\Omega_{d-1}^2)$, with $\zeta\equiv\cos\theta$ and Euclidean time $\tau\equiv it$. For $d>1$, the solution regular at $\theta=0$ is~\cite[eqs.~(144)--(145)]{IW04}
\begin{equation}
u_{\omega J}=(\sin\theta)^J\zeta^{\Delta_+}{}_2F_1\Big(\frac{J+\Delta_+-\omega}{2},\frac{J+\Delta_++\omega}{2};J+\tfrac d2;\sin^2\theta\Big),
\label{eq:uomega}
\end{equation}
with source coefficient $A=\Gamma(J+\frac d2)\Gamma(\nu)/\prod_\pm\Gamma[(J+\Delta_+\pm\omega)/2]$ and response~\cite[eq.~(154)]{IW04}
\begin{equation}
\mathcal{R}_J(\omega)\equiv\frac{B}{A}=\frac{\Gamma(-\nu)}{\Gamma(\nu)}\prod_\pm\frac{\Gamma[(J+\Delta_+\pm\omega)/2]}{\Gamma[(J+\Delta_-\pm\omega)/2]}.
\label{eq:RJ}
\end{equation}
For $d=1$, the same expressions with $J=0,1$ give the even and odd strip solutions, respectively.
For $0<\nu<1$, the counterterm in eq.~\eqref{eq:action} is $S_{\rm ct}\equiv\frac{\Delta_-}{2}\int\!\sqrt{\gamma}\,\Phi^2$, where $\gamma$ is the induced metric on the cutoff surface. Finite source counterterms are set to zero, as in eq.~\eqref{eq:G_R}.

\paragraph{Normal modes.}Before imposing the Robin condition, we differentiate the radial equation with respect to $\omega$, integrate by parts, and evaluate the resulting boundary term using the two asymptotic fall-offs to obtain 
\begin{equation}
2\omega\int_0^{\pi/2}\!\mathrm{d}\theta\,\tan^{d-1}\theta\;u_{\omega J}^2=2\nu A^2\mathcal{R}_J'(\omega).
\label{eq:norm}
\end{equation}
Normal modes satisfy $\mathcal{R}_J(\omega_{nJ})=\kappa$. For a positive spatial operator, unit radial modes and unit spherical harmonics give
\begin{equation}
G_R(\tau,\mathbf n,\mathbf n';\kappa)
=\sum_{n,J,\ell}w_{nJ}e^{-\omega_{nJ}|\tau|}
Y_{J\ell}(\mathbf n)Y_{J\ell}^*(\mathbf n'),
\label{eq:Robinmodeexpansion}
\end{equation}
where $\ell$ labels the harmonic degeneracy and $J$ runs over the retained parity sector. The boundary weights $w_{nJ}=[2\nu\mathcal{R}'_J(\omega_{nJ})]^{-1}$ are minus the residues of $\widehat G_R(\omega,J;\kappa)$. Figure~\ref{fig:pot} shows rescaled representatives of the modes.

The $d=1$ strip has two boundaries, and $J=0,1$ label even and odd extensions across its centre, with responses $\mathcal{R}_+\equiv\mathcal{R}_0$ and $\mathcal{R}_-\equiv\mathcal{R}_1$. For an odd field with one independent endpoint ($A_L=-A_R$), the full-strip norm and boundary action are twice their half-strip values. The endpoint source and weight are therefore $j_R=4\nu(\kappa A_R-B_R)$ and $w_n=[4\nu\mathcal{R}_-'(\omega_n)]^{-1}$. A unit-normalised odd $S^0$ harmonic instead has the $2\nu$ normalisation of eq.~\eqref{eq:G_R}. Projection onto even circle harmonics in $d=2$ does not double a harmonic norm.

\paragraph{Domain and stability.}The Neumann frequencies are $\omega^{N}_{nJ}=a+J+2n$. The conditions $0<\nu<1$ and $a+J>0$ in every retained harmonic give the stationary covariance in eq.~\eqref{eq:Da}. At nonzero Robin coupling, self-adjointness alone does not ensure positivity. Between Dirichlet poles, $\partial_{\omega^2}\mathcal R_J>0$. In the lowest retained sector, the lowest mode reaches $\omega=0$ at $\kappa=\mathcal R_J(0)$, the boundary of the nonnegative range for $d>1$~\cite[theorem~3.2(ii), eq.~(169)]{IW04}. For $\kappa>\mathcal R_J(0)$, the spatial operator is positive and the Fock vacuum is stable; at threshold the zero mode has no stationary vacuum.

\paragraph{Boundary symmetries.}For a boundary conformal Killing vector with $\mathcal L_\xi g_{(0)}=2\sigma_\xi g_{(0)}$, the coefficients $A$ and $B$ carry weights $\Delta_-$ and $\Delta_+$. At constant Robin coupling, their transformations give
\begin{equation}
\left.\delta_\xi(B-\kappa A)\right|_{B=\kappa A}
=-2\nu\kappa\,\sigma_\xi A.
\label{eq:broken}
\end{equation}
Thus $\nu\kappa\neq0$ preserves only the cylinder isometries ($\sigma_\xi=0$), time translations and rotations, and breaks the remaining $2d$ generators. These generators remain symmetries of the bulk equation, but they do not preserve the Robin boundary condition.

For the $d=1$ strip, reflection $\mathsf P:\theta\mapsto-\theta$ also reverses the noncompact generators: $\mathsf P\mathsf E_\pm\mathsf P^{-1}=-\mathsf E_\pm$, as follows from the global AdS$_2$ generators in Ref.~\cite[eq.~(63)]{HSB21}. They therefore exchange even and odd fields. Fixed pair parity restricts the symmetry even at Neumann.

%% file: appendices/C_syk.tex
\section{SYK reduction and the exceptional sector}
\label{app:syk}
\subsection{Seed and boundary normalisation}
To fix the Robin coupling, we reduce the zero-rung term in the microscopic conventions of section~\ref{sec:syk}, with $q>2$ and $\Delta=1/q$. Write $T_{ij}\equiv T_i-T_j$ for microscopic time differences. Maldacena and Stanford's zero-rung term and conformal kernel are~\cite{MS16}
\begin{align}
C_0(T_1,\ldots,T_4)&=-G_c(T_{13})G_c(T_{24})+G_c(T_{14})G_c(T_{23}),\label{eq:SYKseed}\\
K_c(T_1,T_2;T_3,T_4)&\equiv-(q-1)J^2G_c(T_{13})G_c(T_{24})G_c(T_{34})^{q-2}.\label{eq:SYKkernel}
\end{align}
In eq.~\eqref{eq:Fg}, $v_0=(q-1)J^2b^{q-2}/2$, and $K=C_0\mathcal V=K_c$ as operators on antisymmetric bilocals with the full ordered-pair measure; the two exchange terms cancel the $1/2$ in $v_0$. The coupling and amplitude obey $J^2b^q=(\tfrac12-\Delta)\tan(\pi\Delta)/\pi$, so $\alpha_0\equiv[(q-1)J^2b^q]^{-1}=2\pi q/[(q-1)(q-2)\tan(\pi/q)]$.

Splitting the chain integral at $z=0,T_{12}$ gives
\begin{equation}
\int\mathrm{d}z\,G_c(z)G_c(z-T_{12})
=b^2\big[2\mathrm{B}(\nu,1-2\nu)-\mathrm{B}(\nu,\nu)\big]|T_{12}|^{-2a},
\label{eq:sykhF}
\end{equation}
with $\mathrm{B}$ the Euler beta function, initially for $1/4<\Delta<1/2$. For $q\ge4$ ($\Delta\le1/4$), the two Wick-exchange centre integrals diverge separately in the infrared, but their difference converges; we combine them before integrating and continue in $\Delta$. At $q=4$ ($a=0$), taking the prefactor and the difference of powers together gives the finite logarithmic limit $C_0^{(0)}=2b^2\log|(x-y)/(x+y)|$. Using $K=C_0\mathcal V$, we compare the resulting cylinder kernel with eq.~\eqref{eq:Da} to obtain $c_\partial=-2\nu\,4^{-\nu}(q-1)$ and hence the Robin coupling in eq.~\eqref{eq:cSYK}. With this normalisation, eq.~\eqref{eq:kR} becomes $k_c(h)\,\mathcal R_-(h-\tfrac12)=4^{-\nu}(q-1)$, where $k_c(h)$ is the conformal kernel eigenvalue of Ref.~\cite{MS16}.

\subsection{Nonzero reduced families}
\label{app:sykope}
We show that the ordinary SYK families have nonzero centre reductions. In the conventions of Ref.~\cite{MS16}, $\mathcal F_g(\chi)\equiv F_g/[G_c(x)G_c(y)]$, where $\chi\equiv T_{12}T_{34}/(T_{13}T_{24})$. The normalised four-point correction is $\mathcal F_g/N$. We define
\begin{equation}
\begin{gathered}
\mu(h)\equiv\frac{h-\tfrac12}{\pi\tan(\pi h/2)},\qquad
\mathcal N(h)\equiv\frac{\Gamma(h)^2}{2\Gamma(2h)},\\
\mathcal B_h(\chi)\equiv\chi^h{}_2F_1(h,h;2h;\chi).
\end{gathered}
\label{eq:SYKOPEconventions}
\end{equation}
For an ordinary pole, let $\mathbb B_h$ denote its contribution with unit leading OPE coefficient. We take $\mathbb B_h\equiv\mathcal B_h$ for $0<\chi<1$ and $\mathbb B_h\equiv\Psi_h/[2\mathcal N(h)]$ for $\chi>1$, where $\Psi_h$ is the SYK partial wave. These are the two residue representations of Ref.~\cite[eqs.~(3.50), (3.52)]{MS16}. The negative-$\chi$ region follows from bilocal exchange, $\mathbb B_h(\chi)=\mathbb B_h(\chi/(\chi-1))$. We use this prescription throughout the centre integral.

For $x,y>0$, put $\tau\equiv\log(x/y)$ and $u\equiv Y/\sqrt{xy}$. After factoring out $G_c(x)G_c(y)$, the centre integral $\mathcal I$ of appendix~\ref{app:A} takes the dimensionless form $\mathcal P$, with
\begin{equation}
F_g^{(0)}(x,y)=b^2(xy)^{-a}(\mathcal P\mathcal F_g)(\tau).
\label{eq:SYKcentreRescaling}
\end{equation}
We evaluate the centre integral by first introducing an auxiliary parameter $\eta>1$, so that the conformal block remains in $0<\chi<1$. With $\omega\equiv h-\tfrac12$ and $\operatorname{Re}h>\tfrac12$, direct integration gives
\begin{equation}
\mathcal J_h(\eta)
\equiv\int_{-\infty}^{\infty}\!\mathrm du\,
\mathcal B_h\!\left(\frac{2}{1+\eta+2u^2}\right)
=\frac{\pi}{2\omega\mathcal N(h)}
e^{-\omega\operatorname{arcosh}\eta}.
\label{eq:SYKauxiliaryIntegral}
\end{equation}
We then continue $\eta$ to $-\cosh\tau\pm i0$. On the principal branch, $\operatorname{arcosh}(-\cosh\tau\pm i0)=|\tau|\pm i\pi$, giving
\begin{equation}
\mathcal J_h(-\cosh\tau\pm i0)
=\frac{\pi}{2\omega\mathcal N(h)}
e^{-\omega|\tau|}e^{\mp i\pi\omega}.
\label{eq:SYKcontinuedIntegral}
\end{equation}
The piecewise prescription for $\mathbb B_h$ combines these two analytic continuations with weights $[1\pm i\tan(\pi h/2)]/2$. Its centre integral is therefore
\begin{equation}
\begin{aligned}
(\mathcal P\mathbb B_h)(\tau)
&\equiv\int_{-\infty}^{\infty}\!\mathrm du\,
\mathbb B_h\!\left(\frac{1}{u^2-\sinh^2(\tau/2)}\right)\\
&=\frac12\left[
\left(1+i\tan\frac{\pi h}{2}\right)
\mathcal J_h(-\cosh\tau+i0)\right.\\[-2pt]
&\hspace{3.2em}\left.+
\left(1-i\tan\frac{\pi h}{2}\right)
\mathcal J_h(-\cosh\tau-i0)\right]\\
&=\mathcal I(h)e^{-(h-\frac12)|\tau|},\qquad
\mathcal I(h)\equiv\frac{1}{2\mu(h)\mathcal N(h)}.
\end{aligned}
\label{eq:SYKfamilyprojection}
\end{equation}
For $\tau\ne0$, the boundary limits introduce only integrable logarithmic singularities, while the integrand decays as $|u|^{-2\operatorname{Re}h}$ at infinity. The continuation therefore gives the full centre integral for $\operatorname{Re}h>\tfrac12$, away from poles of the family function.
Here $\mathcal I(h)$ is the scalar factor multiplying the reduced family. Since $v_0b^2=1/(2\alpha_0)$, eq.~\eqref{eq:main} gives $\mathcal P\mathcal F_g=2c_\partial\alpha_0G_R$.

For the physical poles $k_c(h_n)=1/g$, the known kernel~\cite[eq.~(3.35)]{MS16}, together with monotonicity of the Robin response, gives $2n+1+2\Delta<h_n(g)<2n+2$ for $0<g<1$ and $n=0,1,\ldots$. At $g=0$, the lower endpoint is reached. At $g=1$, only the lowest root reaches $h=2$; the higher roots remain inside these intervals. Thus every ordinary family has a finite nonzero $\mathcal I(h_n)$ and a distinct positive frequency. Their reduced exponentials are therefore linearly independent. The exceptional $h=2$ contribution is treated in appendix~\ref{app:csyk}.

For the infinite tower, we must also justify reducing the sum term by term. At fixed $g$, the known OPE residues in Refs.~\cite[eq.~(3.31)]{GRline} and~\cite[eq.~(3.54)]{MS16} give coefficients that grow polynomially in the basis $2\mathcal N(h)\mathbb B_h$, whose functions satisfy the Casimir equation across the cuts and decay exponentially with $h$ outside them. For $\tau>0$, the centre-integration weight is smooth at the cuts, so repeated integration by parts controls the sum there, while the remaining endpoint and tail integrals converge. The reduction therefore commutes with summation, and appendix~\ref{app:independence} gives uniqueness of the resulting absolutely convergent exponential expansion.

\subsection{Conformal SYK and the \texorpdfstring{$h=2$}{h=2} sector}
\label{app:csyk}
In the normalisation of Ref.~\cite{GRline}, conformal SYK has a nonlocal kinetic term and the large-$N$ propagator $G(T)=\bar b\,\mathrm{sgn}(T)/|T|^{2\Delta}$, with $\bar b=1/2$ at zero coupling. Rescaling $\psi$ by $(b/\bar b)^{1/2}$ fixes the rails to $G_c$. Combining the kernel multiplier of Ref.~\cite[eq.~(3.25)]{GRline} with our Robin identification gives
\begin{equation}
g=1-2\bar b,\qquad \kappa=g\,4^{-\nu}(q-1),\qquad 0\le g<1.
\label{eq:cSYK}
\end{equation}
Multiplying the covariance by $(\bar b/b)^2$ restores the original fields. The scalar mass is unchanged: varying $g$ changes the Robin coupling on the same AdS$_2$ strip.

For the $h=2$ contribution, we keep $g$ variable during the centre reduction and take $g\to1$ only afterwards. With the rails fixed, the regulated conformal ladder of Ref.~\cite[eq.~(3.45) and footnote~5]{MS16} has spectral factor $k_c(h)/[1-gk_c(h)]$. For $0<g<1$, let $h_0(g)$ be its lowest OPE pole, with $k_c(h_0)=1/g$ and $h_0\to2$ as $g\to1$. Using the OPE residue of Ref.~\cite[eq.~(3.31)]{GRline} in our fixed-rail normalisation, eq.~\eqref{eq:SYKfamilyprojection} gives
\begin{equation}
\begin{aligned}
\mathcal P\!\left[Nc_0^2(g)\mathbb B_{h_0(g)}\right]
&=\frac{\alpha_0}{g^2k_c'(h_0(g))}
e^{-(h_0(g)-\frac12)|\tau|}\\
&\xrightarrow{g\uparrow1}\frac{\alpha_0}{k_c'(2)}e^{-3|\tau|/2}.
\end{aligned}
\label{eq:SYKh2reducedlimit}
\end{equation}
This finite limit is the Robin contribution at $\omega=3/2$.

The unreduced ladder contains the divergent discrete $h=2$ term of Ref.~\cite[eq.~(3.45)]{MS16}. We may therefore define a finite unreduced representative by subtracting this term:
\begin{equation}
\mathcal{F}_1^\perp(\chi)\equiv\lim_{g\uparrow1}\Big[\mathcal{F}_g(\chi)-\frac{3\alpha_0}{\pi^2(1-g)}\,\Psi_2(\chi)\Big].
\label{eq:h2finite}
\end{equation}
At $h=2$, $\mathbb B_2=6\Psi_2$ and the centre integral remains absolutely convergent. Equation~\eqref{eq:SYKfamilyprojection} gives $\mathcal I(2)=0$ and hence $\mathcal P\Psi_2=0$, so
\begin{equation}
\mathcal P\mathcal F_1^\perp=\lim_{g\uparrow1}\mathcal P\mathcal F_g.
\label{eq:SYKsubtractionReduction}
\end{equation}
The subtraction therefore leaves the reduced answer unchanged. The $\omega=3/2$ mode is the reduction of the finite double-pole residue of this regulated conformal ladder~\cite[eqs.~(3.50), (3.52)]{MS16}. In the full finite-coupling SYK four-point function, this residue must be combined with the other $h=2$ corrections~\cite[section~III.C.4]{MS16}.

%% file: appendices/D_ar.tex
\section{AR reduction and normalisation}
\label{app:ar}
\paragraph{Conventions.}
We retain the field notation of Ref.~\cite{BD20} used in section~\ref{sec:ar}. Indices in eq.~\eqref{eq:AR} are raised and lowered with $g_j^{mm'}=g^j_{mm'}\equiv(-1)^{j-m}\delta_{m,-m'}$, with $(\phi^m)^*=\phi_m$; $\lambda_2$ is tuned to criticality (zero in analytic regularisation).

\paragraph{Ladder normalisation.}
The cubic Wick contractions give $\Sigma=\lambda^2G^2/2$, as in Ref.~\cite[eq.~(3.1)]{BD20}. With $G=\mathcal Z_{\rm LR}C$, the rescaled field $\phi^m/\sqrt{\mathcal Z_{\rm LR}}$ has propagator $C$, cubic coupling $\lambda\mathcal Z_{\rm LR}^{3/2}$, and kinetic coefficient $\mathcal Z_{\rm LR}$. The physical rung strength is the square of the cubic coupling, $g\equiv\lambda^2\mathcal Z_{\rm LR}^3$. We use
\begin{equation}
C(x)\equiv\int\frac{\mathrm{d}^2p}{(2\pi)^2}\,e^{ip\cdot x}C(p)=b|x|^{-4/3},
\qquad b\equiv\frac{\Gamma(\tfrac23)}{2^{2/3}\pi\Gamma(\tfrac13)}.
\label{eq:ARpropagatorNormalisation}
\end{equation}
For the rescaled fields, the seed is $C_0\equiv C_{13}C_{24}+C_{14}C_{23}$. Differentiating the self-energy on symmetric bilocals fixes $K=C_0(C_{34}/2)$. At cubic coupling $\lambda\mathcal Z_{\rm LR}^{3/2}$, the action-derived eigenvalue obeys $g\,k(h,J)=\tfrac12k_{1/3}(h,J)$, where $k_{d/6}$ is the expression printed in Ref.~\cite[eq.~(6.19)]{BD20}. Thus the action and Schwinger--Dyson equation give half the printed eigenvalue. We use this action-derived normalisation throughout.

The chain integral fixes $c_\partial=1/(2\pi)$, hence $\kappa=-3g/(8\pi)$ and $k(h,J)=-3/[8\pi\mathcal R_J(h-1)]$ for the retained even spins $J=0,2,\ldots$.

\subsection{Invertibility of the centre reduction}
\label{app:arope}
The centre reduction can be inverted directly. Define $\mathcal F_g^{\rm AR}\equiv F_g/[C(x)C(y)]$, identify $x,y,Y$ in eq.~\eqref{eq:F0} with complex numbers, and write $x/y=e^{\tau+i\varphi}$. The cross-ratio is $z\equiv xy/[Y^2-(x-y)^2/4]$. Set
\begin{equation}
u\equiv\frac{Y}{\sqrt{xy}},\qquad
\xi\equiv\frac{(x-y)^2}{4xy}=\sinh^2\!\frac{\tau+i\varphi}{2},\qquad
f_g(t)\equiv\mathcal F_g^{\rm AR}(1/t,1/\bar t).
\label{eq:ARinversecoordinates}
\end{equation}
The two-to-one change of variables $t\equiv u^2-\xi=1/z$ gives, with $G_R$ regarded as a function of $\xi$,
\begin{equation}
\begin{aligned}
G_R(\xi)=\frac{\pi b^3}{2}\int_{\mathbb C} \mathrm d^2t, \, \frac{f_g(t)}{|\xi+t|}, \qquad f_g(-\xi)&=\frac{1}{\pi^2b^3}(-\Delta_\xi)^{1/2}G_R(\xi).
\end{aligned}
\label{eq:ARinverse}
\end{equation}
Here $\Delta_\xi$ is the flat Laplacian in the $\xi$-plane. The inverse follows because $|\xi|^{-1}$ has Fourier transform $2\pi/|\mathbf k|$, while $(-\Delta_\xi)^{1/2}$ multiplies by $|\mathbf k|$.

The inverse holds distributionally when $\int_{\mathbb C}|f_g(t)|/(1+|t|)\,\mathrm d^2t<\infty$. Ordinary physical blocks satisfy this condition: $h>1$ gives a $|t|^{-h}$ tail, and the singularities at $t=0,1$ are at most squared logarithms. For the full positive ladder with $0\leq g<g_c$, the $L^1$ norms of the reduced rung terms on the cylinder form a geometric series with ratio $g/g_c$. Its centre integral therefore converges almost everywhere, and $f_g$ satisfies the same weighted bound. Positive-time cylinder data cover $\mathbb C\setminus[-1,0]$ and determine the locally integrable image almost everywhere. Thus eq.~\eqref{eq:ARinverse} recovers the full four-point function and its OPE coefficients. It also ensures that the reductions of nonzero independent blocks remain nonzero and independent.

%% file: appendices/E_dgjs.tex
\section{Map to the DGJS interval}
\label{app:dgjs}
We map the DGJS equation and boundary conditions to our scalar. For SYK, take $q>2$, $\Delta=1/q$ and $\nu=1-2\Delta$. Let $y\in(0,1)$ parametrise the positive half of the interval of Ref.~\cite{DGJS17a}. Its even modes obey~\cite[eqs.~(3.4), (3.8), (3.11)]{DGJS17a}
\begin{equation}
\Big[y(1-y)\partial_y^2+\big(2\Delta-(\tfrac12+2\Delta)y\big)\partial_y-\big(\Delta-\tfrac14\big)^2\Big]f_k=-\frac{k^2}{4}f_k,
\label{eq:DGJSode}
\end{equation}
with measure $\mathrm{d}y\,y^{2\Delta-1}/(2\sqrt{1-y})$. On $0<x<\pi/2$, the substitution $y\equiv\cos^2x$, $u_k\equiv(\cos x)^{2\Delta-1/2}f_k(\cos^2x)$ preserves the norms and gives
\begin{equation}
\Big[-\partial_x^2+\frac{(1-2\Delta)^2-\frac14}{\cos^2x}\Big]u_k=k^2u_k,
\label{eq:DGJSradial}
\end{equation}
the stationary equation of our scalar with $k=\omega=h-\tfrac12$. With $f_k=A+By^\nu+\cdots$ at $y=0$, the defect condition $B=(V/2)\,A$ is our Robin condition. Gamma duplication turns the coupling of Ref.~\cite[eq.~(3.18)]{DGJS17a} into
\begin{equation}
\frac V2=(q-1)\frac{\Gamma(\frac32-\Delta)\Gamma(1-\Delta)\Gamma(2\Delta)}{\Gamma(\frac12+\Delta)\Gamma(2-2\Delta)\Gamma(\Delta)}=4^{-\nu}(q-1).
\label{eq:DGJScoupling}
\end{equation}
This is $\kappa$ at $g=1$. The Dirichlet condition $f_k(1)=0$ gives $u_k(0)=0$. Odd extension across the centre gives the full strip. The map $y=\cos^2x$ folds the two radial halves, mapping both AdS boundaries to $y=0$ and the centre to $y=1$. The even parity of Ref.~\cite{DGJS17a} and our odd strip parity are different reflections.

%% file: appendices/F_folded_string.tex
\section{Map to folded-string in \texorpdfstring{AdS$_2$}{AdS2}}
\label{app:foldedstring}
For the $q>2$ SYK scalar of section~\ref{sec:syk}, write $\Phi(t,x)=e^{-i\omega t}u_\omega(x)$. Using the coordinate map of Ref.~\cite[sec.~2.3]{Vegh24}, we define the wavefunction on the folded-string momentum-fraction interval $z\in(0,1)$ by
\begin{equation}
z\equiv\frac{1+\sin x}{2},\qquad
\varphi_\omega(z)\equiv\sqrt{\frac{2}{\cos x}}\,u_\omega(x).
\label{eq:foldedstringmap}
\end{equation}
The prefactor is chosen to preserve the radial $L^2$ norm,
\begin{equation}
\int_0^1\!\mathrm{d}z\,|\varphi_\omega(z)|^2
=\int_{-\pi/2}^{\pi/2}\!\mathrm{d}x\,|u_\omega(x)|^2.
\label{eq:foldedstringnorm}
\end{equation}
This maps the scalar radial equation to the folded-string interval equation in Ref.~\cite[eq.~(2.4)]{TV26}, with eigenvalue $h(h-1)=\omega^2-\tfrac14$. Near $z=0$, $\varphi_\omega\sim c_Az^{-\nu/2}+c_Bz^{\nu/2}$, so $B=\kappa A$ becomes $c_B=4^\nu\kappa c_A$, with the same coefficient relation at $z=1$. At the SYK coupling in eq.~\eqref{eq:SYKdata}, this gives $c_B=(q-1)c_A$, the fractional-power Robin condition of Ref.~\cite[eqs.~(2.12)--(2.13)]{TV26}. Odd strip modes obey $\varphi_\omega(1-z)=-\varphi_\omega(z)$.

%% file: references.tex
\bibliographystyle{JHEP}
\bibliography{references}